\documentclass[12pt]{article}
\usepackage{amsmath,amsfonts,epsfig}
\numberwithin{equation}{section}

\ifx\ltimes\undefined
\newcommand{\ltimes}{{\kern3pt\hbox{\vrule width 0.4pt height 5.30pt
depth .0pt}\kern-1.76pt\times\kern1pt}} \fi

\ifx\lrtimes\undefined
\newcommand{\rtimes}{{\kern1pt\times\kern-4.76pt\kern3pt\hbox{\vrule width 0.4pt height 5.30pt
depth .0pt}}} \fi

\def\Z {\mathbb{Z}}
\def\R {\mathbb{R}}

\def\bid{\hbox{1\hspace{-0.04in}I}} 

\def\d{\delta}

\def\m{\mu}
\def\n{\nu}
  
\def\s{\sigma}                                   

\def\G{\Gamma}

\def\cX{{\cal X}}

\begin{document}

\begin{titlepage}
\begin{flushleft}
\hfill  Imperial/TP/06/RAR/02 \\
\hfill  QMUL-PH-06-04\\
\hfill  hep-th/0603094 \\

\end{flushleft}
\vspace*{8mm}

\begin{center}

{\Large {Flux Compactifications of\\ M-Theory on Twisted
Tori}} \\

\vspace*{12mm}

{ C.M.~Hull$^{1,2}$ and R.A.~Reid-Edwards$^{1,3}$
} \\
\vspace*{7mm} {\em $^1$The Institute for Mathematical Sciences}\\
{\em Imperial College London} \\
{\em 48 Prince's Gardens, London SW7 2PE, U.K.} \\

\vspace*{4mm}

{\em $^2$The Blackett Laboratory, Imperial College London} \\
{\em Prince Consort Road, London SW7 2AZ, U.K.} \\

\vspace*{4mm}

{\em $^3$ Department of Physics}\\
{\em Queen Mary, University of London}\\
{\em Mile End Road, London, E1 4NS, U.K.}

\vspace*{12mm}

\end{center}

\begin{abstract}

We find the bosonic sector of the gauged supergravities that are obtained from 11-dimensional supergravity by Scherk-Schwarz dimensional reduction with flux to any dimension $D$.
We show that, if certain obstructions are absent, the  Scherk-Schwarz ansatz for a finite set of $D$-dimensional fields can be extended
to a full compactification of M-theory, including an infinite tower of Kaluza-Klein fields.
The internal  space   is obtained from a group manifold (which may be non-compact)
by a discrete identification.
We discuss the symmetry algebra and the symmetry breaking patterns and illustrate these with particular examples.
We discuss the action of U-duality on  these theories in terms of symmetries of the $D$-dimensional supergravity, and argue that in general it will take geometric flux compactifications to M-theory on non-geometric backgrounds, such as U-folds with U-duality transition functions.

\end{abstract}

\vfill

\noindent {Email: { c.hull@imperial.ac.uk}, {
r.reid-edwards@imperial.ac.uk} }

\end{titlepage}

\section{Introduction}

In \cite{Scherk:1979zr}, Scherk and Schwarz gave an ansatz for a
non-trivial dimensional reduction of a supergravity theory that
gives a theory with gauge symmetry, mass terms and a scalar
potential. The dimensional reduction
 from $D+d$ dimensions is on a $d$ dimensional internal manifold ${\cal X}$ with
a basis of nowhere-vanishing one-forms $\sigma^m$ specified by a vielbein $\sigma^m{}_i(y)$
\begin{equation} \sigma^m= \sigma^m{}_i (y) dy^i
\end{equation}
where $y^i$ are coordinates on   ${\cal X}$ ($i,j=1,...,d$, $m,n=1,...,d$). The one-forms are used in the metric ansatz
\begin{equation}\label{SS metric}
ds_{D+d}^2=e^{2\alpha\varphi(x)}ds^2_D+e^{2\beta\varphi(x)}g_{mn}(x)\nu^m\nu^n
\end{equation}
where
   $M$ is the $D$-dimensional spacetime with metric $ds^2_D$ and coordinates $x^\m$. The one-forms
$\nu^m$ are
\begin{equation}
\nu^m=\sigma^m-A^m
\end{equation}
and the one-forms $A^m=A^m_\mu (x)dx^\mu $ are Kaluza-Klein vector
fields  (graviphotons). The internal metric $g_{mn}(x)$ and the
warp factor $\varphi(x)$ are scalars in the dimensionally reduced
theory, while $\alpha, \beta $ are constants that will be fixed in section 2. There is a similar ansatz for the reduction of other
fields. For example, for a $p$-form gauge potential
\begin{equation}
\label{B-field ansatz}
\widehat{B}_{(p)}=B_{(p)}+B_{(p-1)m}\wedge\nu^m+\frac{1}{2!}B_{(p-2)mn}\wedge
\nu^m\wedge\nu^n+...
\end{equation}

Completeness of the basis implies that the one-forms satisfy a structure equation of the form
\begin{equation}
\label{eq:maurer-cartan} d\sigma^m+\frac{1}{2}f_{np}{}^m\sigma^n
\wedge \sigma^p=0
\end{equation}
If the coefficients $f_{np}{}^m$ are   {\it constant}, then   the dependence of the dimensionally reduced theory on  the internal coordinates
$y$ drops out \cite{Scherk:1979zr}. The integrability condition for (\ref{eq:maurer-cartan}) is then the Jacobi identity
\begin{equation}
\label{eq:ff=0} f_{[mn}{}^qf_{p]q}{}^t=0
\end{equation}
so that $f_{np}{}^m$ are the structure constants for some Lie group $G$, and the reduced theory has a local $G$ gauge symmetry for which the
gauge fields are the $A^m$. For a reduction of the action to be possible, it is necessary \cite{Scherk:1979zr} that the structure constants
further satisfy
\begin{equation}
\label{eq:f=0} f_{mn}{}^n=0
\end{equation}
If this condition is not satisfied, then it is possible to reduce
the field equation, even though it may not be possible to reduce
the action \cite{Bergshoeff:1997mg}.

The internal space ${\cal X}$ is sometimes referred to as a twisted torus, and the matrix $ \sigma ^m{}_i (y)$ can be thought of as   twisting
the frames with respect to the coordinate basis. Indeed, the fields $g_{\m\n}, A^m_\m, g_{mn}, \varphi$ and $B_{\m_1...\m_p},
B_{\m_1...\m_{p-1}m}, B_{\m_1...\m_{p-2}mn}....$ etc are in one-to-one correspondence with the ones that would arise from reduction on a
$d$-torus, and the reduced theory is a massive deformation of that arising from a torus reduction. Particular examples of such reductions can
arise from first reducing on $T^{d-1}$ and then reducing on the final $S^1$ with a geometric duality twist \cite{Hull:2005hk}. In such examples
this is equivalent to the reduction of a torus bundle over a circle \cite{Hull:1998vy}, which can be thought of as a twisting of the torus
$T^d$. However, an important class of examples are those in which $G$ is compact and the internal manifold is the group manifold $G$ and it is
clearly misleading to refer to such a group manifold as a twisted torus. There are more general examples in which $G$ is non-compact, but in
which the Scherk-Schwarz ansatz nonetheless gives a consistent truncation to a well-defined $D$-dimensional field theory.

The Scherk-Schwarz construction gives a truncation of a $D+d$ dimensional field theory
in a particular background to give a $D$-dimensional field theory with a finite number of fields. A long-standing question
has been how to extend this to the full Kaluza-Klein or string  theory. For the dimensional reduction of a $D+d$ dimensional field theory, one
would expect towers of massive Kaluza-Klein modes and one would like to know how to calculate the spectrum and whether there is a mass gap.  One
way of obtaining the Scherk-Schwarz reduction is to reduce on the group manifold $G$ and then truncate to the $y$-independent sector. This would
give the same $D$ dimensional field theory discussed above, and in the case in which $G$ is compact gives a compactification of the original
theory. However, if $G$ is non-compact, then  if one were to include the $y$-dependence one would expect a continuous mass spectrum in general.
A well-behaved Kaluza-Klein theory with a discrete mass spectrum would be obtained if there was a compact space ${\cal X}$  such that
compactification on ${\cal X}$ could be truncated  to reproduce the Scherk-Schwarz reduction. This would then allow the construction to be
extended to compactification of string theory or M-theory on $\cX$.

The issue of finding such a $\cX$ and hence understanding the Scherk-Schwarz reduction as a compactification was addressed in
\cite{Hull:2005hk}. If $G$ is compact, one simply takes $\cX$ to be the group manifold with the $\s$ the left-invariant forms for which
(\ref{eq:maurer-cartan}) is the Maurer-Cartan equation. There is a left-action $G_L$ and a right-action $G_R$ of the group $G$ on the group
manifold. The Scherk-Schwarz ansatz is the most general one that is invariant under $G_L$ so that the Scherk-Schwarz reduction is a
compactification followed by a truncation to a $G_L$-invariant sector. The metric (\ref{SS metric}) will not be invariant under $G_R$ unless
$g_{mn}$ is chosen to be an invariant metric (otherwise the background will break $G_R$ to the subgroup preserving $g_{mn}$).
 The one-forms
$\nu^m$ are invariant under $G_L$ but transform covariantly under
a local ($x^\m$-dependent) action of $G_R$, which becomes a gauge
symmetry in the compactified theory, with the frame indices $m,n$
becoming adjoint gauge indices, so that e.g. $g_{mn}$ transforms
in the symmetric bi-adjoint representation of the gauge group $G$.

For non-compact $G$, one requires a compact space $\cX$ with frame fields satisfying (\ref{eq:maurer-cartan}). Then  ${\cal X}$ must be locally
isomorphic to the group manifold $G$, but this need not be true globally. Moreover,  the consistency of the ansatz requires that the frame
fields should be globally defined and nowhere-vanishing, so that $\cX$ must be parallelizable. (For example, suppose that the $\s^m$ are
sections of the frame bundle of the internal space, so that in overlaps of patches on $\cX$ in which the coordinates $y,y'$ are related by a
diffeomorphism $y'(y)$, then $\s '{}^m (y'(y)) = \Lambda ^m{}_n (y) \s^n(y)$ for some local frame rotation $ \Lambda ^m{}_n (y)$ and the
corresponding internal metrics would need to be related by $g_{mn} = g'{}_{pq}  \Lambda ^p{}_m  \Lambda ^q{}_n$, which would not allow them to
be indpendent of $y$.) This implies that the internal manifold must either be the group manifold itself, or the group manifold identified under
the action of a  discrete subgroup $\Gamma$  of $G_L$ \cite{Hull:2005hk}. It is important that $\G$ acts as a subgroup of $G_L$, so that the
one-forms $\s$ are invariant under $\G$ and the local form of the ansatz takes the same form on $G$ or $G/\G$. Thus for a compactification to be
possible, one requires that there exist a discrete subgroup $\G$ of $G$ such that identifying the group manifold under the left action of $\G$
gives a compact space $\cX=G/\G$ which can be taken as the compactifying space \cite{Hull:2005hk}. A necessary condition for the existence of
such a $\G$ is that the structure constants satisfy the condition (\ref{eq:f=0}).

If there is a compact space $\cX=G/\G$, then supergravity, string theory or  M-theory can be compactified on $\cX$ in the usual way. The
Scherk-Schwarz construction is then a consistent truncation of the full compactified theory to a $D$-dimensional effective field theory with
fields $g_{\m\n}, A^m_\m, g_{mn}, \varphi$ and $B_{\m_1...\m_p}, B_{\m_1...\m_{p-1}m}, B_{\m_1...\m_{p-2}mn}....$ etc. Note that this truncated
set may not contain all the light fields in general. For example, if $G$ is compact and $\G$ is trivial, then at the special point in moduli
space in which $g_{mn}$ is proportional to the Cartan-Killing metric and all the form gauge fields vanish, the background has isometry
$G_L\times G_R$ which will be a gauge symmetry in the reduced theory, so that  there will be $2d$ massless Yang-Mills gauge fields. The
Scherk-Schwarz construction truncates this theory to a $G_L$-invariant sector of the low-energy theory with only $d$ gauge fields and gauge
symmetry $G_R$.

In this paper, we will consider the compactification of M-theory
on $d$-dimensional
 twisted tori ${\cal X}=G/\G$
 with flux.
 This has a truncation to a Scherk-Schwarz reduction of
eleven dimensional supergravity   \cite{Cremmer:1978km}
 in which the ansatz for the 3-form gauge field  $\widehat{C}$ is generalised to include
 a flux for the  field
strength $\widehat{G}=d\widehat{C}$ of the form
\begin{equation}\label{flux}
{\cal K}= \frac{1}{24}K_{mnpq} \s^m\wedge \s^n\wedge \s^p\wedge \s^q
\end{equation}
for some constant coefficients $K_{mnpq}$; such a flux is manifestly invariant under $G_L$. This generalises the compactification of string
theory on twisted tori
 with flux \cite{Hull:2005hk} that truncate to
 generalised Scherk-Schwarz reductions with flux \cite{Hull:2005hk,Odd,Cvetic:2003jy}.
 Such reductions of 11-dimensional supergravity have also been considered in \cite{D'Auria:2005er,Dall'Agata:2005mj,D'Auria:2005dd,Dall'Agata:2005ff,Dall'Agata:2005fm,D'Auria:2005rv,Fre:2005px,Fre':2006ut}.

The $O(d,d)$ covariant formulation of string theory reduced on a twisted torus, studied in \cite{Odd},
\cite{Hull:2005hk} is very suggestive. It is natural
to ask whether this generalises to   M-theory compactifications, and whether these can be written in a way that is covariant under the action of
a duality group. We consider general Scherk-Schwarz compactifications of eleven dimensional supergravity with flux and analyse the gauge
symmetry. In this case the generators $X^m$ related to the $B_{(1)m}$ fields in \cite{Hull:2005hk} are replaced by generators $X^{mn}=- X^{nm}$
which can be associated with the field $C_{(1)mn}$ which arises from the dimensional reduction of the three form potential $\widehat{C}$ of the
eleven dimensional theory.

The outline of the paper is as follows: In section two we give the Scherk-Schwarz reduction of eleven dimensional
supergravity, where there is a non-trivial flux on the four form, to arbitrary dimensions. In section three we study the symmetries of this
theory, in particular we show that the symmetry algebra of this theory is not a Lie algebra in general, but contains a Lie subalgebra. Section
four deals with symmetry breaking and mass mechanisms in such reductions and then finally in section five we consider the writing of these
theories in a manifestly $E_{d(d)}$-covariant form. In section six we discuss the implications of our results for M-Theory.

\section{Scherk-Schwarz Reduction of Eleven Dimensional Supergravity with Flux}

The action of  eleven dimensional supergravity is
\begin{equation}
{\cal S}=\int{\cal L}_B+{\cal L}_F
\end{equation}
where the Lagrangian for the bosonic sector is
\begin{equation}
\label{cjs lagrangian}{\cal L}_{B}=\widehat{R}* 1 - \frac{1}{2}
*\widehat{G} \wedge\widehat{G} + \frac{1}{6}\widehat{G} \wedge
\widehat{G} \wedge \widehat{C}
\end{equation}
and the four-form field strength $\widehat{G}_{(4)}$ is defined in
terms of a three form potential $\widehat{C}_{(3)}$
\begin{equation}
\widehat{G} = d \widehat{C}
\end{equation}
${\cal L}_F$ is the Fermi sector involving the gravitino $\widehat{\psi}_{\mu}$. In this paper the fermions are set to zero and we consider a
Scherk-Schwarz reduction with flux of the bosonic sector following \cite{Hull:2005hk,Odd,Cvetic:2003jy}.

We adopt the metric ansatz (\ref{SS metric}) and the   $G_L$-invariant
flux ansatz
\begin{equation}
\widehat{G}=\frac{1}{24}K_{mnpq}\sigma^m\wedge\sigma^n\wedge\sigma^p\wedge\sigma^q+...
\end{equation}
for constant $K_{mnpq}$. We require that the constants $K_{mnpq}$
satisfy the algebraic identity
\begin{equation}
\label{c-field integrability}K_{[mnp|s}f_{|qt]}{}^s=0
\end{equation}
so that the flux  is closed and so locally there is a 3-form $\varpi_{(3)}$ such that
\begin{equation}
d\varpi_{(3)}=\frac{1}{24}K_{mnpq}\sigma^m\wedge\sigma^n\wedge\sigma^p\wedge\sigma^q
\end{equation}
In general $\varpi_{(3)}$ is not defined globally. Then the $G_L$-invariant reduction ansatz for the three-form is
\begin{equation}
\label{c-field ansatz
}\widehat{C}=C_{(3)}+C_{(2)m}\wedge\nu^m+\frac{1}{2}C_{(1)mn}\wedge\nu^m\wedge\nu^n+\frac{1}{6}C_{(0)mnp}\nu^m\wedge\nu^n\wedge\nu^p+\varpi_{(3)}
\end{equation}
The field strength $\widehat{G}=d\widehat{C}$ is
\begin{eqnarray}
\widehat{G}&=&G_{(4)}+G_{(3)m}\wedge\nu^m+\frac{1}{2}G_{(2)mn}\wedge\nu^m\wedge\nu^n+\frac{1}{6}G_{(1)mnp}\wedge\nu^m\wedge\nu^n\wedge\nu^p\nonumber\\
&&+\frac{1}{24}G_{(0)mnpq}\nu^m\wedge\nu^n\wedge\nu^p\wedge\nu^q
\end{eqnarray}
where the reduced field strengths are
\begin{eqnarray}
\label{eq:c-field strengths}G_{(4)}&=&dC_{(3)} - C_{(2)m}\wedge
F^m - \frac{1}{24}K_{mnpq}A^m\wedge A^n\wedge A^p\wedge A^q
\nonumber\\
G_{(3)m}&=&DC_{(2)m} - C_{(1)mn}\wedge
F^n+\frac{1}{6}K_{mnpq}A^n\wedge A^p\wedge A^q
\nonumber\\
G_{(2)mn}&=&DC_{(1)mn} - C_{(2)p}f_{mn}{}^p - C_{(0)mnp}F^p -
\frac{1}{2}K_{mnpq}A^p\wedge A^q
\nonumber\\
G_{(1)mnp}&=&DC_{(0)mnp} - 3C_{(1)[m|q}f_{|np]}{}^q+K_{mnpq}A^q
\nonumber\\
G_{(0)mnpq}&=&-6C_{(0)[mn|t}f_{|pq]}{}^t - K_{mnpq}
\end{eqnarray}
We note the appearance of Chern-Simons-type terms arising from the
flux. This will have important consequences for the gauge algebra
of the reduced theory as we shall discuss in the following
sections. The $G_R$-covariant derivatives are
\begin{eqnarray}
DC_{(2)m}&=&dC_{(2)m}+f_{mp}{}^nC_{(2)n}\wedge A^p
\nonumber\\
DC_{(1)mn}&=&dC_{(1)mn} + 2f_{[m|q}{}^pC_{(1)|n]p}\wedge A^q
\nonumber\\
DC_{(0)mnp}&=&dC_{(0)mnp}+3f_{[m|t}{}^qC_{(0)|np]q}A^t
\end{eqnarray}
The zero-curvature equations $G_{(p)}=0$ define a Free Differential Algebra \cite{Dall'Agata:2005mj,D'Auria:2005dd,D'Auria:2005rv,Fre:2005px}.
The Bianchi identity $d\widehat{G}=0$ gives the  set of identities
\begin{eqnarray}
dG_{(4)}+G_{(3)m}\wedge F^m=0\nonumber\\
DG_{(3)m}+G_{(2)mn}\wedge F^n=0\nonumber\\
DG_{(2)mn}+G_{(1)mnp}\wedge F^p=0\nonumber\\
DG_{(1)mnp}+G_{(0)mnpq} F^q=0\nonumber\\
DG_{(0)mnpq}=0
\end{eqnarray}

We now have the necessary information to give the reduction to $D$ dimensions of the Lagrangian (\ref{cjs lagrangian}) on the $d$ dimensional
twisted torus ${\cal X}$ with metric ansatz (\ref{SS metric}) and 3-form ansatz (\ref{c-field ansatz }). Up to boundary terms, the bosonic Lagrangian reduced to $D$-dimensions is
\begin{equation}\label{reduced Lagrangian}
{\cal L}_D={\cal L}_\mathcal{R}+{\cal L}_{\widehat{G}}+{\cal
L}^{cs}_D+V*1
\end{equation}
where ${\cal L}_\mathcal{R}$ arises from the reduction of the
eleven dimensional Einstein-Hilbert Lagrangian
\begin{equation}
{\cal L}_\mathcal{R}=R*1 - \frac{1}{2}*d\varphi\wedge d\varphi -
\frac{1}{2}g^{mp}g^{nq}*Dg_{mn}\wedge Dg_{pq}
-\frac{1}{2}e^{2(\beta-\alpha)\varphi}g_{mn}*F^m \wedge F^n
\end{equation}
where $Dg_{mn}=dg_{mn}+g_{mp}f_{nq}{}^pA^q+g_{np}f_{mq}{}^pA^q$ and
\begin{equation}
\alpha=-\left(\frac{d}{2(D-2)(D+d-2)}\right)^{\frac{1}{2}}  \qquad \beta=\left(\frac{D-2}{2d(D+d-2)}\right)^{\frac{1}{2}}
\end{equation}
have been chosen to give the dilaton kinetic term the canonical normalisation and ensure the Lagrangian has an Einstein-Hilbert term without any
conformal prefactors. The reduction of the four form field strength kinetic term gives
\begin{eqnarray}
{\cal L}_{\widehat{G}}&=&-
\frac{1}{2}e^{-4\alpha\varphi}*G_{(4)}\wedge
G_{(4)}-\frac{1}{2}e^{-2(\beta+\alpha)\varphi}g^{mn}*G_{(3)m}\wedge
G_{(3)n}
\nonumber\\
&&-\frac{1}{2}e^{-4\beta\varphi}g^{mn}g^{pq}*G_{(2)mp}\wedge
G_{(2)nq}
\nonumber\\
&&-\frac{1}{2}e^{-2(3\beta-\alpha)\varphi}g^{mn}g^{pq}g^{ts}*G_{(1)mpt}\wedge
G_{(1)nqs}
\end{eqnarray}
The Scherk-Schwarz reduction generates a potential $V$ in the
effective theory where
\begin{eqnarray}
V&=&-\frac{1}{4}e^{2(\alpha-\beta)\varphi}\left(
g_{mn}g^{pq}g^{ts}f_{pt}{}^mf_{qs}{}^n+2g^{mn}f_{qm}{}^pf_{pn}{}^q\right)
\nonumber\\
&&-\frac{1}{2}e^{-4(2\beta-\alpha)\varphi}g^{mn}g^{pq}g^{ts}g^{lj}G_{(0)mptl}
G_{(0)nqsj}
\end{eqnarray}
Both the geometry and the flux contribute to the potential. The
${\cal L}^{cs}_D$ are dimension dependent terms arising from the
reduction of the eleven dimensional Chern-Simons term
\begin{equation}
{\cal L}^{cs}_{D+d}=\frac{1}{6}\widehat{G}\wedge \widehat{G}\wedge
\widehat{C}
\end{equation}
to $D$-dimensions. The reduction of the Chern-Simons term is given explicitly in the Appendix and generalises
that of \cite{Lu:1995yn} to include flux. Such reductions to $D=4$ have been considered in \cite{D'Auria:2005er,Dall'Agata:2005mj}.

\section{Gauge Symmetry Algebra}

In this section we consider the gauge symmetries of the reduced
Lagrangian. The gauge group arises from anti-symmetric tensor
transformations and diffeomorphisms on the twisted torus. The
field strengths are invariant under the infinitesimal
 anti-symmetric tensor transformations
$\widehat{C}\rightarrow \widehat{C}+d\widehat{\lambda}$, where
\begin{eqnarray}\label{lambda}
\widehat{\lambda}&=&\Omega_{(2)}+\Lambda_{(1)m}\wedge\nu^m+\frac{1}{2}\lambda_{(0)mn}\nu^m\wedge\nu^n
\nonumber\\
d\widehat{\lambda}&=&\left(d\Omega_{(2)}+\Lambda_{(1)m}\wedge
F^m\right)+\left(D\Lambda_{(1)m}+\lambda_{(0)mn}F^n\right)\wedge
\nu^m
\nonumber\\
&&+\frac{1}{2}\left(\Lambda_{(1)p}f_{mn}{}^p+D\lambda_{(0)mn}
\right)\wedge\nu^m\wedge\nu^n+\frac{1}{2}\lambda_{(0)mq}f_{np}{}^q\nu^m\wedge\nu^n\wedge\nu^p
\nonumber\\
\end{eqnarray}
and the reduced parameters $\lambda_{(0)mn}$, $\Lambda_{(1)m}$,
and $\Omega_{(2)}$ are   the parameters of independent scalar,
one-form and two-form anti-symmetric tensor transformations
respectively. The corresponding $D$-dimensional gauge
transformations of the reduced potentials are
\begin{eqnarray}
\label{c-field gauge}
\delta_X(\widehat{\lambda})C_{(3)}&=&d\Omega_{(2)}+\Lambda_{(1)m}\wedge
F^m \nonumber\\
\delta_X(\widehat{\lambda})C_{(2)m}&=&D\Lambda_{(1)m}+\lambda_{(0)mn}F^n
 \nonumber\\
\delta_X(\widehat{\lambda})C_{(1)mn}&=&\Lambda_{(1)p}f_{mn}{}^p+D\lambda_{(0)mn}
\nonumber\\
\delta_X(\widehat{\lambda})C_{(0)mnp}&=&3\lambda_{(0)[m|q}f_{|np]}{}^q
\end{eqnarray}
Diffeomorphisms on the internal manifold lead to a second set of Yang-Mills gauge transformations with parameter $\omega^m(x)$. The requirement
that $\widehat{C}$ is invariant under general coordinate transformations
and the covariant transformation of the one-form $\nu^m$
\begin{equation}
\delta(\omega)\nu^m=-\nu^nf_{np}{}^m\omega^p
\end{equation}
induces the following transformations on the reduced potentials
\begin{eqnarray}
\delta(\omega)C_{(3)}&=&\frac{1}{6}K_{mnpq}\omega^qA^m\wedge
A^n\wedge A^p+d\Xi_{(2)}+\Xi_{(1)m}\wedge F^m \nonumber\\
\delta(\omega)C_{(2)m}&=&C_{(2)n}f_{mp}{}^n\omega^p+\frac{1}{2}K_{mnpq}\omega^qA^n\wedge
A^p+D\Xi_{(1)m}+\Xi_{(0)mn}F^n \nonumber\\
\delta(\omega)C_{(1)mn}&=&2C_{(1)[m|p}f_{|n]q}{}^p\omega^q+K_{mnpq}\omega^qA^p+\Xi_{(1)p}f_{mn}{}^p+D\Xi_{(0)mn}
\nonumber\\
\delta(\omega)C_{(0)mnp}&=&3C_{(0)[mn|q}f_{|p]t}{}^q\omega^t+K_{mnpq}\omega^q+3\Xi_{(0)[m|q}f_{|np]}{}^q
\end{eqnarray}
where $\widehat{\Xi}\equiv\iota_{\omega}\varpi_{(3)}$ and
\begin{equation}
\widehat{\Xi} =\Xi_{(2)}+\Xi_{(1)m}\wedge\nu^m+\frac{1}{2}\Xi_{(0)mn}\nu^m\wedge\nu^n
\end{equation}
has explicit dependence on the internal coordinates\footnote{In
general, $\widehat{\Xi}$ will not be left-invariant but the
Lagrangian is still invariant under this transformation.} of
$\mathcal{X}$. We remove this dependence  on the internal
coordinates by a gauge transformation $\widehat{C}\rightarrow
\widehat{C}+d\widehat{\lambda}$ with parameter
$\widehat{\lambda}=-\widehat{\Xi}(y)$ yielding the infinitesimal
gauge transformations
\begin{eqnarray}
\label{c-field algebra } \delta_Z(\omega)A^m&=&-D\omega^m
\nonumber\\
\delta_Z(\omega)C_{(3)}&=&\frac{1}{6}K_{mnpq}\omega^qA^m\wedge
A^n\wedge A^p \nonumber\\
\delta_Z(\omega)C_{(2)m}&=&C_{(2)n}f_{mp}{}^n\omega^p+\frac{1}{2}K_{mnpq}\omega^qA^n\wedge
A^p \nonumber\\
\delta_Z(\omega)C_{(1)mn}&=&2C_{(1)[m|p}f_{|n]q}{}^p\omega^q+K_{mnpq}\omega^qA^p
\nonumber\\
\delta_Z(\omega)C_{(0)mnp}&=&3C_{(0)[mn|q}f_{|p]t}{}^q\omega^t+K_{mnpq}\omega^q
\end{eqnarray}
We have included the transformation of the Kaluza-Klein vector
fields $A^m$.

The gauge symmetries
$(\delta_Z(\omega),\delta_X(\widehat{\lambda}))$ generate the
gauge algebra
\begin{eqnarray}\label{c-field algebra}
\lbrack
\delta_Z\left(\tilde{\omega}^m\right),\delta_Z\left(\omega^n\right)
\rbrack &=&\delta_Z\left(f_{np}{}^m\omega^n
\tilde{\omega}^p\right) - \delta_X\left(K_{mnpq}\omega^p
\tilde{\omega}^q\right)
\nonumber\\
&&-\delta_W\left(K_{mnpq}\omega^p \tilde{\omega}^qA^n\right) -
\delta_\Sigma\left(\frac{1}{2}K_{mnpq}\omega^p
\tilde{\omega}^qA^m\wedge A^n\right)
\nonumber\\
\lbrack \delta_X\left(
\lambda_{(0)mn}\right),\delta_Z\left(\omega^p\right) \rbrack
&=&\delta_X\left(\lambda_{(0)mp}f_{nq}{}^p\omega^q\right) -
\delta_X\left(\lambda_{(0)np}f_{mq}{}^p\omega^q\right)
\nonumber\\
\lbrack \delta_W\left(
\Lambda_{(1)m}\right),\delta_Z\left(\omega^p\right) \rbrack
&=&\delta_W\left(\Lambda_{(1)n}f_{mp}{}^n\omega^p\right)
\end{eqnarray}
where the antisymmetric tensor transformations $\delta_X(\widehat{\lambda})$ has been split
  into $\delta_X(\lambda_{(0)mn})$, $\delta_W(\Lambda_{(1)m})$ and $\delta_{\Sigma}(\Omega)$. All other commutators vanish and the identities
$f_{[mn}{}^qf_{p]q}{}^t=0$ and $K_{[mnp|s}f_{|qt]}{}^s=0$ have
been used. With a little work, again using the identities
$f_{[mn}{}^qf_{p]q}{}^t=0$ and $K_{[mnp|s}f_{|qt]}{}^s=0$, it can
be checked that this algebra satisfies the Jacobi identity
\begin{equation}
[[\delta_A(\alpha),\delta_B(\beta)],\delta_C(\gamma)]+[[\delta_B(\beta),\delta_C(\gamma)],\delta_A(\alpha)]
+[[\delta_C(\gamma),\delta_A(\alpha)],\delta_B(\beta)]=0
\end{equation}
where $\delta_A(\alpha)$, $\delta_B(\beta)$ and $\delta_C(\gamma)$
denote any of $\delta_Z(\omega^m)$, $\delta_X(\lambda_{(0)mn})$,
$\delta_W(\Lambda_{(1)m})$ and $\delta_{\Sigma}(\Omega_{(2)})$.

\subsection{Field dependent parameters and Chern-Simons terms}

We shall briefly comment on the field dependent terms in the gauge
algebra, analogous to those found in \cite{Hull:2005hk} for the
Kalb-Ramond field. These terms, surprising at first, arise
generically in theories with field strengths that include
Chern-Simons-like terms such as those in the previous section. As
an illustration, consider the simpler case of a three-form field
strength
\begin{equation}
H_{(3)}=dB_{(2)}-{\cal Q}_{(3)}
\end{equation}
where
\begin{equation}
{\cal Q}_{(3)}=tr\left(A\wedge dA+\frac{1}{3}A\wedge A\wedge
A\right)
\end{equation}
is a Chern-Simons term satisfying $d{\cal Q}_{(3)}=tr(F\wedge F)$ where $F$ is the two-form field strength $F=dA+A\wedge A$. The one-form $A $
transforms as a Yang-Mills connection $\delta_Z (\epsilon) A=-D\epsilon$. The requirement that $H_{(3)}$ be gauge invariant means $B_{(2)}$, in
addition to the antisymmetric tensor transformation $\delta_X(\lambda)B_{(2)}=d\lambda_{(1)}$, must transform under $\delta_Z(\epsilon)$ as
$\delta_Z (\epsilon)B_{(2)}=\epsilon dA$. The gauge algebra realised on $B_{(2)}$ is then
\begin{equation}
[\delta_Z(\epsilon),\delta_Z(\tilde{\epsilon})]=\delta_X
(\epsilon\tilde{\epsilon}A)
\end{equation}
which has a field dependent parameter. This is a specific example
of a more general phenomenon involving Chern-Simons terms.

As an example consider the field strength $G_{(3)m}$ with
$f_{mn}{}^p=0$ and $K_{mnpq}\neq 0$. In this case we may write
\begin{eqnarray}
G_{(3)m}&=&dC_{(2)m}-{\cal Q}_{(3)m}\nonumber\\
{\cal Q}_{(3)m}&=&C_{(1)mn}\wedge F^n-\frac{1}{6}K_{mnpq}A^n\wedge
A^p\wedge A^q
\end{eqnarray}
where $d{\cal Q}_{(3)m}=G_{(2)mn}\wedge F^n$. This leads to the field dependent algebra (\ref{c-field algebra}).

\subsection{Lie Subalgebra}

The Yang-Mills  gauge transformations are $\delta_Z(\omega)$, $\delta_X(\lambda_{(0)})$ and comparing with the algebra (\ref{c-field algebra})
and the discussion of \cite{Odd} suggests that this might correspond to a gauge group with Lie algebra   of the form
\begin{eqnarray}
\label{field algebra
3}\left[Z_m,Z_n\right]&=&-f_{mn}{}^pZ_p+K_{mnpq}{X}^{pq}
\nonumber\\
\left[{X}^{mn},Z_p\right]&=&-f_{pq}{}^m{X}^{nq}+f_{pq}{}^n{X}^{mq}
\nonumber\\
\left[{X}^{mn},{X}^{pq}\right]&=&0
\end{eqnarray}
where $Z_m$ and $X^{mn}$ are the group generators for the
transformations $\delta_Z(\omega)$ and  $\delta_X(\lambda_{(0)})$
respectively. The extra field-dependent terms in the algebra
(\ref{c-field algebra}) would then arise from the
Chern-Simons-like terms, as discussed in section 3.1. This would
be the direct analogue of the string case discussed in
\cite{Hull:2005hk}. However, the algebra (\ref{field algebra 3})
does not satisfy the Jacobi identities and so is not a Lie
algebra, so the situation cannot be so straightforward.

To understand the gauge algebra further, consider gauge
transformations with parameter
\begin{equation}
\lambda_{(0)mn}=\frac{1}{2}f_{mn}{}^p\breve{\lambda}_{(0)p}
\end{equation}
From (\ref{c-field gauge}), the effect of the
$\breve{\lambda}_{(0)m}$ transformation can all be absorbed in a
redefinition
\begin{equation}
\breve{\Lambda}_{(1)m}= \Lambda_{(1)m}+ D \breve{\lambda}_{(0)m}
\end{equation}
so that transformations of this form do not act. This is because, gauge fields $C_{(1)mn}$ of the form
\begin{equation}
C_{(1)mn}=\frac{1}{2}f_{mn}{}^p\breve{C}_{(1)p}
\end{equation}
can be absorbed into a field redefinition
\begin{equation}
\breve{C}_{(2)m}=  C_{(2)m}- D \breve{C}_{(1)m}
\end{equation}
so that $ C_{(2)m}$ becomes massive by \lq eating' $\breve{C}_{(1)m}$, the gauge boson of the $\breve{\lambda}_{(0)m}$ transformation, so the
gauge symmetry with parameter $\lambda_{(0)mn}=\frac{1}{2}f_{mn}{}^p\breve{\lambda}_{(0)p}$ is broken by any vacuum of the theory. The remaining
gauge fields are the $\breve{C}_{(1)mn}$ that are in some sense orthogonal to the gauge fields
$C_{(1)mn}=\frac{1}{2}f_{mn}{}^p\breve{C}_{(1)p}$. For semi-simple groups, these can be defined by taking them to be orthogonal to the
$\breve{C}_{(1)p}$ with respect to the Cartan-Killing metric, while the definition  for general groups will be postponed until section 4.

The gauge generators ${X}^{mn}$ can now be decomposed into a part $\breve{X}^{mn}$ satisfying $f_{mn}{}^p\breve{X}^{mn}=0$ and a part
$X^p=f_{mn}{}^p\breve{X}^{mn}$ generating the transformations with parameter $\lambda_{(0)mn}=\frac{1}{2}f_{mn}{}^p\breve{\lambda}_{(0)p}$. As
we have seen, the $\breve{X}^p$ transformations are broken by a choice of vacuum, leaving the algebra of $Z_m$ and $\breve{X}^{mn}$
transformations given by
\begin{eqnarray} \label{c-field algebra
3}\left[Z_m,Z_n\right]&=&-f_{mn}{}^pZ_p+K_{mnpq}\breve{X}^{pq}
\nonumber\\
\left[\breve{X}^{mn},Z_p\right]&=&-f_{pq}{}^m\breve{X}^{nq}+f_{pq}{}^n\breve{X}^{mq}
\nonumber\\
\left[\breve{X}^{mn},\breve{X}^{pq}\right]&=&0
\end{eqnarray}
In this case, the Jacobi identity holds identically
\begin{equation}
[[Z_m,Z_n],Z_p]+[[Z_p,Z_m],Z_n]+[[Z_n,Z_p],Z_m]=K_{mnpq}f_{ts}{}^q\breve{X}^{ts}=0
\end{equation}
by virtue of the condition $f_{mn}{}^p\breve{X}^{mn}=0$ and this is a Lie sub-algebra of the full symmetry group.

\section{Symmetry Breaking and Examples of Flux Reductions}

The reduction on a twisted torus with flux gives rise to a compactified theory with the gauge algebra (\ref{c-field algebra}). This symmetry
will in general be spontaneously broken by any given vacuum of the theory. First, some of the gauge symmetry is non-linearly realised, and as a
non-linearly realised transformation acts as a shift on certain fields $\phi$, $\d \phi = \alpha + O(\phi)$, it cannot be preserved by any
vacuum expectation value of $\phi$ and so is necessarily broken by any vacuum, so that the gauge group is necessarily broken down to its
linearly realised subgroup. Then any given vacuum solution (e.g. one
 arising from   a critical point of the scalar potential) can
then  break the linearly realised subgroup further to the subgroup
preserving that vacuum.

In this section, we will discuss the first stage of symmetry
breaking down to the linearly realised subgroup that is generic
for any solution. For vacua with vanishing scalar expectation
value, this is the complete breaking, but for non-trivial scalar
expectation values there will be further breaking through the
standard Higgs mechanism. The transformation for the scalar fields
$C_{(0)mnp}$ is
\begin{equation}
\delta C_{(0)mnp}=3\lambda_{(0)[m|q}f_{|np]}{}^q+K_{mnpq}\omega^q +O(C_{(0)mnp})
\end{equation}
and from this one can read off the non-linearly realised symmetries, i.e. the ones realised as shifts of scalar fields. The non-linear
transformation of the $C_{(1)mn}$ fields occurs in a similar way
\begin{equation}
\delta C_{(1)mn}=\Lambda_{(1)p}f_{mn}{}^p+O(C_{(1)mn})
\end{equation}

\subsection{Trivial Flux}

Consider the flux
\begin{equation}
K_{mnpq}=\zeta_{mnt}f_{pq}{}^t - \zeta_{mpt}f_{nq}{}^t+\zeta_{mqt}f_{np}{}^t - \zeta_{nqt}f_{mp}{}^t+\zeta_{npt}f_{mq}{}^t -
\zeta_{qpt}f_{mn}{}^t
\end{equation}
Where $\zeta_{mnp}=\zeta_{[mnp]}$. The effect of this flux is removed by the field redefinition
\begin{eqnarray}
C_{(3)}&\rightarrow&C_{(3)}+\frac{1}{6}\zeta_{mnp}A^m\wedge
A^n\wedge A^p
\nonumber\\
C_{(2)m}&\rightarrow&C_{(2)m}+\frac{1}{2}\zeta_{mnp}A^n\wedge A^p
\nonumber\\
C_{(1)mn}&\rightarrow&C_{(1)mn}+\zeta_{mnp}A^p
\nonumber\\
C_{(0)mnp}&\rightarrow&C_{(0)mnp}+\zeta_{mnp}
\end{eqnarray}
This flux is therefore physically trivial, and any such flux
produces physics that is equivalent to that of a model without
flux.

\subsection{Reduction With Semi-Simple Group $G$}

Consider the reduction of M-Theory on a twisted torus ${\cal X}\simeq G/\Gamma$ where $G$ is a, not necessarily compact, semi-simple group, and
$\G\subset G_L$ is a discrete subgroup such that $G/\G$ is compact. The Scherk-Schwarz reduction on such a twisted torus produces a theory in
which the $C_{(2)m}$, $C_{(1)mn}$ and $C_{(0)mnp}$ fields all become massive through the Higgs mechanism. For example, the term in the reduced
Lagrangian responsible for the $C_{(1)mn}$ field mass is
\begin{equation}
{\cal L}=-\frac{9}{2}e^{2(3\beta-\alpha)\bar{\varphi}}\bar{g}^{mq}\bar{g}^{nt}\bar{g}^{ps}C_{(1)[m|l}f_{|np]}{}^lC_{(1)[q|h}f_{|ts]}{}^h+...
\end{equation}
where $\bar{g}$ and $\bar{\varphi}$ are vacuum values for the scalar fields. In addition to the appearance of Goldstone scalars $\chi_{(0)mn}$
for the broken symmetries with scalar parameter $\lambda_{(0)mn}$, the symmetry breaking requires a set of Goldstone one-forms $\chi_{(1)m}$
corresponding to the breaking of the gauge symmetries with parameter $\Lambda_{(1)m}$, a feature that is qualitatively distinct from the
analysis of string theory discussed in \cite{Hull:2005hk}, but is generic for higher degree forms. For a semi-simple group the Cartan-Killing
metric $\eta_{mn}$, defined by
\begin{equation}
\eta_{mn}=-\frac{1}{2}f_{mp}{}^qf_{nq}{}^p
\end{equation}
is non-degenerate and invertible. The inverse Cartan-Killing metric $\eta^{mn}$ may be used to raise the indices of the structure constants
$f_m{}^{np}=\eta^{nq}f_{mq}{}^p$. If the flux is zero, the full gauge algebra of the theory (\ref{c-field algebra}) corresponds to the Lie
algebra
\begin{eqnarray}
\left[Z_m,Z_n\right]&=&-f_{mn}{}^pZ_p \nonumber\\
\left[X^{mn},Z_p\right]&=&-f_{pq}{}^mX^{nq}+f_{pq}{}^nX^{mq}
\nonumber\\ \left[W^{m},Z_n\right]&=&-f_{np}{}^mW^p
\end{eqnarray}
with all other commutators vanishing. For convenience, the constants $O^{qt}_{mnp}$ and $\Pi^{mnp}_{qt}$ are defined\footnote{Various useful
identities that these constants satisfy may be found in the Appendix.}
\begin{eqnarray}
\label{eq:O definition}O^{qt}_{mnp}&=&3\delta^q{}_{[m}f_{np]}{}^t
\nonumber\\ \Pi^{mnp}_{qt}&=&\frac{1}{2}\delta^{[m}{}_qf_t{}^{np]}
\end{eqnarray}
These constants will be seen to play an analogous role in the three-form symmetry breaking mechanism to $f_{mn}{}^p$ and $f_p{}^{mn}$ in
\cite{Hull:2005hk}. The fields which become massive are singlets of the antisymmetric tensor transformations generated by
$\delta_X(\lambda_{(0)mn})$ and $\delta_X(\Lambda_{(1)m})$. The gauge transformation generated by $\delta_X(\Omega_{(2)})$ is not charged under
the right action $G_R$ acting on $\cX$ and therefore plays no role in the symmetry breaking mechanism defined here. This is to be expected since
the only field that is charged under this symmetry is $C_{(3)}$ which has no mass-like term in the Lagrangian and is expected to remain the
massless gauge boson of the $\delta_X(\Omega_{(2)})$ transformation. This mechanism works analogously to the two-form case for the
$\delta_W(\Lambda_{(1)m})$ symmetry, however the $\delta_X(\lambda_{(0)mn})$ symmetry requires more care and it is consideration of this sector
that motivates the introduction of the constants $O_{mnp}^{qt}$ and $\Pi^{mnp}_{qt}$ above. The $\delta_X(\lambda_{(0)mn})$ parameter is
decomposed into irreducible representations of $G_R$
\begin{equation}
\lambda_{(0)mn}=\breve{\lambda}_{(0)mn}+\frac{1}{2}f_{mn}{}^p\breve{\lambda}_{(0)p}
\end{equation}
where $f_p{}^{mn}\breve{\lambda}_{(0)mn}=0$. The $\delta_X(\lambda_{(0)mn})$ gauge transformation is now split into two orthogonal parts
$\delta_X(\breve{\lambda}_{(0)mn})$ and $\delta_X(\breve{\lambda}_{(0)m})$. The potentials transform as
\begin{eqnarray}
\label{eq:stuckelberg gauge 1 }
\delta_X(\lambda_{(0)mn})C_{(2)m}&=&\breve{\lambda}_{(0)mn}F^n+\frac{1}{2}D^2\breve{\lambda}_{(0)m}
\nonumber\\
\delta_X(\lambda_{(0)mn})C_{(1)mn}&=&D\breve{\lambda}_{(0)mn}+\frac{1}{2}f_{mn}{}^pD\breve{\lambda}_{(0)p}
\nonumber\\
\delta_X(\lambda_{(0)mn})C_{(0)mnp}&=&O^{qt}_{mnp}\breve{\lambda}_{(0)qt}
\end{eqnarray}
where the equality $D^2\lambda_{(0)m}=\lambda_{(0)p}f_{mn}{}^pF^n$ has been used. Note that $\breve{\lambda}_{(0)m}$ now only enters into the
gauge transformations as the covariant derivative $D\breve{\lambda}_{(0)m}$, and $O^{qt}_{mnp}f_{qt}{}^s=0$ by virtue of the identity
(\ref{eq:ff=0}) so that $C_{(0)mnp}$ is a singlet of the transformation generated by $\breve{\lambda}_{(0)m}$. The corresponding Goldstone
fields of the broken $\delta_X(\breve{\lambda}_{(0)m})$, $\delta_X(\breve{\lambda}_{(0)mn})$ and $\delta_X(\lambda_{(1)m})$ symmetries are
defined as
\begin{eqnarray}
\label{stuckelberg c-field}
\chi_{(0)mn}&=&\Pi^{pqt}_{mn}C_{(0)pqt} \nonumber\\
\chi_{(1)m}&=&\frac{1}{2}f_m{}^{np}C_{(1)np}
\end{eqnarray}
where $f_p{}^{mn}\chi_{(0)mn}=0$. Using (\ref{eq:stuckelberg gauge
1 }) and (\ref{stuckelberg c-field}) one finds that these fields
transform as\footnote{Note that the definition of $\chi_{(0)mn}$
means that is has general symmetry, in particular
$\chi_{(0)(mn)}\neq 0$.}
\begin{eqnarray}
\delta(\omega,\widehat{\lambda})\chi_{(0)mn}&=&\breve{\lambda}_{(0)mn}+\chi_{(0)mp}f_{nq}{}^p\omega^q+\chi_{(0)pn}f_{mq}{}^p\omega^q
\nonumber\\
\delta(\omega,\widehat{\lambda})\chi_{(1)m}&=&\frac{1}{2}D\breve{\lambda}_{(0)m}+\Lambda_{(1)m}+\chi_{(1)n}f_{mp}{}^n\omega^p
\end{eqnarray}
It is now simple to construct potentials $\breve{C}$ that are invariant under the infinitesimal $\delta_X(\breve{\lambda}_{(0)mn})$,
$\delta_X(\breve{\lambda}_{(0)m})$ and $\delta_X(\Lambda_{(1)m} )$ transformations
\begin{eqnarray}
\breve{C}_{(3)}&=&C_{(3)}-\chi_{(1)m}\wedge F^m \nonumber\\
\breve{C}_{(2)m}&=&C_{(2)m}-\chi_{(0)mn}F^n-D\chi_{(1)m}
\nonumber\\
\breve{C}_{(1)mn}&=&C_{(1)mn}-D\chi_{(0)[mn]}-f_{mn}{}^p\chi_{(1)p}
\nonumber\\
\breve{C}_{(0)mnp}&=&C_{(0)mnp}-O^{qt}_{mnp}\chi_{(0)qt}
\end{eqnarray}
$\breve{C}_{(3)}$ is not a $\delta_X(\Omega_{(2)})$ singlet and as such it remains massless, as expected. (In section 5 we will consider cases
in seven dimensions with a non trivial flux in which $C_{(3)}$ has a topological mass arising form the Chern-Simons term in the Lagrangian.)
Note that the Goldstone boson $\chi_{(1)m}$ for the broken symmetry with parameter $\Lambda_{(1)m}$ is also the gauge boson for the symmetry
with parameter $\breve{\lambda}_{(0)m}$. Thus the gauge boson for the symmetry generated by $\delta_X(\breve{\lambda}_{(0)m})$ is eaten by the
$C_{(2)m}$ fields. This is a general result that extends to reductions on twisted tori ${\cal X}=G/\Gamma$ where $G$ is not semi-simple. We
shall ignore the symmetry $\delta_X(\breve{\lambda}_{(0)m})$ in the following as it is always spontaneously broken in this way. Since these
field redefinitions enter in the same form as gauge transformations the form of the field strengths will be unchanged, except that now we have
massive $\breve{C}$ fields which are singlets under the gauge transformations where we previously had $C$ fields, which transform under the
anti-symmetric gauge transformations. The algebra gauged by the $\breve{C}$ fields is
\begin{equation}\label{broken algebra}
\left[Z_m,Z_n\right]=-f_{mn}{}^pZ_p
\end{equation}
All other commutators vanish, and so the gauge symmetry is broken to the semi-simple group $G_R$ i.e. the algebra generated by $Z_m$ and
$X^{mn}$ is broken to the subalgebra (\ref{broken algebra}) generated by $Z_m$. The symmetry may be broken further by a choice of the constant
metric vacuum expectation $\bar{g}_{mn}$. The gauge group $G_R$ will be broken to the isometry group of the metric $G_{\bar{g}}\subset G_R$, for
which
\begin{equation}
\delta_Z(\omega)\bar{g}_{mn}=2\bar{g}_{(m|p}f_{|n)q}{}^p\omega^q=0
\end{equation}
The Cartan-Killing metric preserves the full gauge group but a more general choice of metric, as is allowed in the Scherk-Schwarz ansatz, will
not giving a reduced theory in which the graviphotons of the broken symmetries become massive with mass
\begin{equation}
{\cal L}_D=-\left(\bar{g}^{mn}\bar{g}^{pq}f_{mt}{}^pf_{ns}{}^q-2\eta_{ts}\right)*A^t\wedge A^s+...
\end{equation}

In the case of a two form $\widehat{B}_{(2)}$ with flux $K_{mnp}$ it was shown ~\cite{Hull:2005hk,Cvetic:2003jy} that one could introduce the
flux $K_{mnp}=f_{mnp}=\eta_{[m|q}f_{|np]}{}^q$, for semi-simple Lie group $G$. It is natural to ask whether there is an analogous form for the
four-form flux, constructed from the structure constants with an ansatz of the form
\begin{equation}\label{flux guess}
K_{mnpq}= \zeta_{[mn|s}f_{|pq]}{}^s
\end{equation}
in which case there might also be similar field redefinitions that could be found explicitly. However, in this case the integrability condition
$K_{[mnp|s}f_{|qt]}{}^s=0$ implies that any flux of the form (\ref{flux guess}) must have a tensor $\zeta_{mnp}=\zeta_{[mnp]}$ in which case the
flux is a trivial flux of the form discussed in section 4.1

\subsection{$T^d$ Reduction with Flux}

If $f_{mn}{}^p=0$, then the group $G_R$ is abelian and the internal manifold (after discrete identifications to compactify, if necessary) is   a
torus. With flux $K$, the gauge Lie algebra (\ref{c-field algebra 3}) for such compactifications is
\begin{equation}
\left[Z_m,Z_n\right]=K_{mnpq}X^{pq}
\end{equation}
with all other commutators vanishing. The inclusion of the flux in the field strength $G_{(1)mnp}=K_{mnpq}A^q+...$ gives a mass-like term for
the graviphotons in the low energy action. For a given vacuum expectation value of the scalars $\bar{g}$ and $\bar{\varphi}$ the graviphoton
mass term is
\begin{equation}
\mathcal{L}_D=-\frac{1}{2}M^2_{lh}*A^{l}\wedge A^{h}+...
\end{equation}
where the mass matrix $M_{mn}$ is given by
\begin{equation}
M^2_{lh}=e^{-\bar{\varphi}}\bar{g}^{mn}\bar{g}^{pq}\bar{g}^{ts}K_{mptl}K_{nqsh}
\end{equation}
The internal index $m$ can be split into $m=(m',\bar{m})$, where $m'=1,2...d'$ and $\bar{m}=d'+1,d'+2...d$ such that, with a suitable choice of
coordinates,
\begin{equation}
K_{mnp\bar{q}} =0 \qquad K_{m'n'p'q'} \neq 0
\end{equation}
Then the transformation of the $C_{(0)mnp}$ scalars is
\begin{equation}
\delta C_{(0)m'n'p'} =K_{m'n'p'q'}\omega^{q'}  \qquad \delta C_{(0)mn\bar{p}} = 0
\end{equation}
The transformations generated by $Z_{m'}$ with parameters $\omega^{m'}$ are spontaneously broken, with $C_{(0)m'n'p'}$ the Goldstone fields that
are eaten by the gauge fields $A^{m'}$. The Lie group is broken to the $d(d+1)/2-d'$ dimensional abelian subgroup $U(1)^{\frac{1}{2}d(d+1)-d'}$
generated by $Z_{\bar m}$ and $X^{mn}$ with parameters $\omega^{\bar{m}}$ and $\lambda_{(0)mn}$ respectively.

Let $\tilde K^{m'n'p'q'}$  be any constants satisfying $\tilde K^{m'n'p'q'}K_{ n'p'q't'}=\delta ^{m'}{}_{t'}$. Then the Goldstone fields
$\chi_{(0)}{}^{m'}$ may be defined by
\begin{equation}
\chi_{(0)}{}^{m'}=\tilde K^{m'n'p'q'}C_{(0)n'p'q'}
\end{equation}
transforming as a shift symmetry. The $\chi_{(0)}{}^{m'}$ transforms as
\begin{equation}
\delta\chi_{(0)}{}^{m'}=\omega^{m'}
\end{equation}
The massive graviphotons $\breve{A}^{m'}=A^{m'}+d\chi_{(0)}{}^{m'}$ may then be defined which are singlets of the gauge transformations.

\subsection{General Case}

In general the group upon which the reduction is based $G$ may be non-semi-simple and the flux will only be constrained to satisfy
$K_{[mnp|s}f_{|qt]}{}^s=0$. To begin, as in section 4.3, the parameter $\lambda_{(0)mn}$ is decomposed into irreducible representations of $G_R$
\begin{equation}
\lambda_{(0)M}=\breve{\lambda}_{(0)M}+\frac{1}{2}f_M{}^m\breve{\lambda}_{(0)m}
\end{equation}
where the compound index\footnote{${d \choose n}=\frac{d!}{n!(d-n)!}$} $\{M\}=\{[mn]\}$ where $M=1,2...{d \choose 2}$ has been used.  Even
though $\breve{\lambda}_{(0)m}$ does not appear as a shift symmetry this symmetry is broken in the effective theory as demonstrated for the
semi-simple case in section 4.2. The transformations of the potentials with shift symmetries are
\begin{eqnarray}
\delta(\omega,\lambda_{(0)})C_{(0)mnp}&=&\breve{\lambda}_{(0)qt}O^{[qt]}_{mnp}+K_{mnpq}\omega^q+...   \nonumber\\
\delta(\omega,\Lambda_{(1)})C_{(1)mn}&=&\Lambda_{(1)p}f_{mn}{}^p+\frac{1}{2}f_M{}^mD\breve{\lambda}_{(0)m}+...
\end{eqnarray}
The breaking of each of the symmetries and the definition of their respective Goldstone bosons are considered in turn. First, consider
transformations of the scalar fields
\begin{eqnarray}
\delta(\omega,\lambda_{(0)})C_{(0)\Sigma}&=&\breve{\lambda}_{(0)M}O^M{}_\Sigma+K_{\Sigma
m}\omega^m+... \nonumber\\
&&=\left(\begin{array}{cc} \breve{\lambda}_{(0)M} & \omega^m
\end{array}\right)\left(\begin{array}{cc}K_{\Sigma
m} \\
O^M{}_\Sigma\end{array}\right)+...\nonumber\\
&&=\alpha_{(0)A}t^A{}_\Sigma+...
\end{eqnarray}
The compound index $\{\Sigma\}=\{[mnp]\}$, $\Sigma=1,2...{d \choose 3}$ has been defined for convenience. The basis $A=1,2,...{d \choose 2}$ is
chosen such that $t^A{}_\Sigma$ takes the form
\begin{equation}
t^A{}_\Sigma=\left(\begin{array}{cc} t^{A'}{}_\Sigma & 0
\end{array} \right)=\left(\begin{array}{cc}t^{A'}{}_{\Sigma'} & 0 \\
0 & 0 \end{array}\right)
\end{equation}
where the split $\{A\}=\{(A',\bar{A})\}$ is defined by the CoKernel and Kernel of the map
\begin{equation}
t:\R^{d \choose 2}\rightarrow \R^{d \choose 3}
\end{equation}
defined by $\alpha_{(0)A}\rightarrow \alpha_{(0)A}t^A{}_\Sigma$. The indices take values $A'=1,2,..d'$ and $\bar{A}=d'+1,..{d \choose 2}$ in the
kernel and cokernel respectively. The choice of the basis $\{\Sigma\}=\{(\Sigma',\bar{\Sigma})\}$ where $\Sigma'=1,2,..d'$ and
$\{\bar{\Sigma}\}=d'+1,..{d \choose 3}$ has also been made. The matrix $t^{A'}{}_{\Sigma'}$ is square and one may define an inverse
$\tilde{t}^{\Sigma'}{}_{A'}$ such that $t^{B'}{}_{\Sigma'}\tilde{t}^{\Sigma'}{}_{A'}=\delta^{B'}{}_{A'}$. It is then possible to define the
Goldstone boson of the broken symmetry, with parameter $\alpha_{(0)A'}$, as
\begin{equation}
\chi_{(0)A'}=C_{(0)\Sigma'}\tilde{t}^{\Sigma'}{}_{A'}
\end{equation}
which transforms as
\begin{equation}
\delta\chi_{(0)A'}=\alpha_{(0)A'}+...
\end{equation}
where the dots denote terms linear in $\chi_{(0)A'}$. The $C_{(0)\Sigma'}$ are eaten by the massive $C_{(1)M'}$ whilst the $C_{(0)\bar{\Sigma}}$
remain as massless scalars, or moduli, of the theory.

Consider the one-form shift symmetry generated by the parameter
$\Lambda_{(1)m}$. The transformation of the $C_{(1)M}$ field may
be written as
\begin{equation}
\delta(\lambda_{(0)m},\Lambda_{(1)m})C_{(1)M}=f_M{}^m\left(\Lambda_{(1)m}+\frac{1}{2}D\breve{\lambda}_{(0)m}\right)+...
\end{equation}
Interpreting $f_M{}^m$ as a map
\begin{equation}
f:\R^d\rightarrow \R^{d\choose 2}
\end{equation}
we choose a basis for the kernel of $f$, labelled by $m'=1,2...d'$ and a basis for the cokernel labelled by $\bar{m}=d'+1,..d$. Then
$\{m\}=\{(m',\bar{m})\}$ and $\{M\}=\{(M',\bar{M})\}$. $f_M{}^m$ may then be written in the form
\begin{equation}
f_M{}^m=\left(\begin{array}{cc} f_M{}^{m'} & 0
\end{array}\right)=\left(\begin{array}{cc} f_{M'}{}^{m'} & 0 \\ 0 & 0
\end{array}\right)
\end{equation}
The matrix $\tilde{f}_{n'}{}^{M'}$ is defined such that $\tilde{f}_{n'}{}^{M'}f_{M'}{}^{m'}=\delta_{n'}{}^{m'}$. The transformations then become
\begin{eqnarray}
\delta(\breve{\lambda}_{(0)m'},\Lambda_{(1)m'})C_{(1)M'}&=&f_{M'}{}^{m'}\left(\Lambda_{(1)m'}
+\frac{1}{2} D
\breve{\lambda}_{(0)m'}\right)+O(C_{(1)M'})\nonumber\\
\delta(\breve{\lambda}_{(0)\bar{m}},\Lambda_{(1)\bar{m}})C_{(1)\bar{M}}&=&O(C_{(1)\bar{M}})
\end{eqnarray}
The symmetries generated by the parameters
$\breve{\lambda}_{(0)m'}$ and $\Lambda_{(1)m'}$ are broken and the
corresponding Goldstone bosons are
\begin{equation}
\chi_{(1)m'}=\tilde{f}_{m'}{}^{M'}C_{(1)M'}
\end{equation}
where
\begin{equation}
\delta\chi_{(1)m'}=\Lambda_{(1)m'}+\frac{1}{2}D\breve{\lambda}_{(0)m'}+...
\end{equation}
The $\breve{C}_{(1)M'}$ are eaten by the $C_{(2)m'}$ which become the massive $\breve{C}_{(2)m'}=C_{(2)m'}-D\chi_{(1)m'}$, whilst the
$C_{(1)\bar{M}}$ and $C_{(2)\bar{m}}$ remain massless.
Various field redefinitions outlined in this section may be performed to bring the algebra to the form
\begin{equation}
\left[{\cal T}_{\bar{M}},{\cal T}_{\bar{N}}\right]=-t_{{\bar{M}{\bar{N}}}}{}^{\bar{P}}{\cal T}_{\bar{P}}+h_{{\bar{M}{\bar{N}}}}{}^a{\cal T}_a
\end{equation}
where the $T_a$ generate a central extension of the unbroken symmetry generated by $T_{\bar{M}}$ (with parameter $\alpha_{(0)}{}^{\bar {M}}$)
and all other commutators vanish. For example, in the case of compactification on the $d+d'$ dimensional twisted torus $G/\G=G_{d}\times T^{d'}$
where $G_d$ is a $d$-dimensional compact semi-simple group manifold, the linearly realised Lie subalgebra is
\begin{equation}
\left[Z_{m},Z_{n}\right]=-f_{mn}{}^{p}Z_{p}+K_{mni}X^{i}
\end{equation}
where $m,n=1,2...d$ label the coordinates $y^m$ on $G_d$ and $i,j=d+1,...d+d'$ label coordinates $y^i$ on the torus $T^{d'}$. In this case the
isometry generators on the torus $Z_i$ and the gauge transformations $X^m$ are always spontaneously broken following arguments similar to those
of the last section.

\section{Duality Covariant Formulations}

In \cite{Hull:2005hk} reductions of a field theory containing gravity, a two-form tensor with flux and a scalar dilaton were studied. The
compact internal manifold was a twisted torus ${\cal X}=G/\G$, where $\G\subset G_L$. It was shown that the lower dimensional theory could be
written in an $O(d,d)$ covariant way where a subgroup $L\subset O(d,d)$ was gauged. This was a truncation of the   results of \cite{Odd} where
the effective low energy field theory of the heterotic string was reduced on a twisted torus with flux. In the heterotic case the reduced
Lagrangian could be written in an $O(d,d+16)$ covariant way. A natural question to ask is whether the general eleven dimensional supergravity
reduction on a twisted torus with flux may be written in a $E_{d(d)}$ covariant form and if so, what is the nature of the interplay between the
global $E_{d(d)}$ group, U-duality and the gauge symmetry. It is this question that we address in this section.

\subsection{String Theory and $O(d,d)$}

First we review the analysis presented in \cite{Hull:2005hk} for the sector consisting of a metric $\widehat{g}$, dilaton $\widehat{\Phi}$ and a
three form field strength which may be written locally in terms of a two form $\widehat{H}_{(3)}=d\widehat{B}_{(2)}$. The low energy Lagrangian
is
\begin{equation}
\label{string frame lagrangian} {\cal
L}_{D+d}=e^{-\widehat{\Phi}}\left( \widehat{R}*1+*d\widehat{\Phi}
\wedge d\widehat{\Phi} - \frac{1}{2}*\widehat{H}_{(3)}\wedge
\widehat{H}_{(3)} \right)
\end{equation}
Using the procedure outlined in section 2 the theory described by this Lagrangian is reduced on a twisted torus with flux
\begin{equation}
{\cal K}=\frac{1}{6}K_{mnp}\sigma^m\wedge\sigma^n\wedge\sigma^p
\end{equation}
for $\widehat{H}_{(3)}$ where we use the Scherk-Schwarz ansatz
\begin{equation}
\widehat{B}=B_{(2)}+B_{(1)m}\wedge\nu^m+\frac{1}{2}B_{(0)mn}\nu^m\wedge\nu^n+\varpi_{(2)}
\end{equation}
where $d\varpi_{(2)}={\cal K}$. The reduced theory may be written in a manifestly $O(d,d)$ covariant way ~\cite{Odd,Hull:2005hk}
\begin{eqnarray}\label{O(d,d) Lagrangian}
{\cal L}_D&=&e^{-\phi}\left(R*1+*d\phi\wedge
d\phi+\frac{1}{2}*G_{(3)}\wedge
G_{(3)}+\frac{1}{4}L_{AC}L_{BD}*D{\cal M}^{AB}\wedge D{\cal
M}^{CD}\right. \nonumber\\ &&- \left.\frac{1}{2}L_{AC}L_{BD}{\cal
M}^{AB}*{\cal F}^C\wedge{\cal F}^D- \frac{1}{12}{\cal M}^{AD}{\cal
M}^{BE}{\cal M}^{CF}t_{ABC}t_{DEF}\right. \nonumber\\ &&\left.+
\frac{1}{4}{\cal M}^{AD}L^{BE}L^{CF}t_{ABC}t_{DEF}\right)
\end{eqnarray}
The scalars parameterise the coset $O(d,d)/O(d)\times O(d)$
\begin{equation} {\cal M}^{AB}= \left(\begin{array}{cc}
g^{mn} & -B_{(0)np}g^{pm} \\ -B_{(0)mp}g^{np} &
g_{mn}+g^{pq}B_{(0)mp}B_{(0)nq}
\end{array}\right)
\end{equation}
and
\begin{eqnarray}\label{Chern-simons field strength}
G_{(3)}&=&d{\cal B}_{(2)}+\frac{1}{2}\left(L_{AB}{\cal A}^A\wedge
{\cal F}^B - \frac{1}{6}t_{ABC}{\cal A}^A\wedge {\cal A}^B\wedge
{\cal A}^C\right) \nonumber\\ {\cal B}_{(2)}&=&B_{(2)} -
\frac{1}{2}B_{(1)m}\wedge A^m \nonumber\\ D{\cal M}^{AB}&=&d{\cal
M}^{AB}+{\cal M}^{AC}t_{CD}{}^B{\cal A}^D+{\cal
M}^{BC}t_{CD}{}^A{\cal A}^D
\end{eqnarray}
where the one-forms fit into an $O(d,d)$ vector ${\cal A}^A$ with field strength ${\cal F}^A$
\begin{equation}{\cal A}^A= \left(\begin{array}{cc}
A^m \\ B_{(1)m}
\end{array}\right)  \qquad  {\cal F}^A= \left(\begin{array}{cc}
F^m \\ G_{(2)m} - B_{(0)mn}F^n
\end{array}\right)
\end{equation}

where
\begin{equation}
G_{(2)m}=DB_{(1)m}+B_{(0)mn}+\frac{1}{2}K_{mnp}A^n\wedge A^p
\end{equation}
Defining $t_{ABC}=L_{AD}t_{BC}{}^D$ where $L_{AB}$
  is the  $O(d,d)$ invariant matrix
  \begin{equation} L_{AB}=\left(\begin{array}{cc}
0 & \bid_d \\ \bid_d & 0
\end{array}\right) \end{equation}
the structure constants are $t _{np}{}^m=f_{np}{}^m$ and $t_{[mnp]}=K_{mnp}$. $\bid_d$ is the d-dimensional identity matrix $\delta_{mn}$.
Upper  indices $m=1,...,d$ indicate covariant vectors under the $GL(d,\R)$ subgroup of $O(d,d)$ while lower indices indicate contravariant vectors.
The
presence of $t_{AB}{}^C$ breaks the $O(d,d)$ symmetry of the ungauged theory to the subgroup preserving $t_{AB}{}^C$. However, the theory
becomes formally invariant under $O(d,d)$ if the constants are taken to transform covariantly under $O(d,d)$. In \cite{Dabholkar:2005ve}, it was
argued that (\ref{O(d,d) Lagrangian}) is the Lagrangian for general gaugings of this sector of the supergravity theory. Some of these gaugings cannot arise from conventional compactifications of supergravity but can
arise from non-geometric compactifications \cite{Dabholkar:2005ve}. In the string theory, $O(d,d)$ is broken to $O(d,d;\Z)$ and this $O(d,d;\Z)$
acts as a T-duality group on the internal space, mixing twist with flux and in general transforming geometric compactifications to non-geometric
ones such as T-folds \cite{Dabholkar:2005ve}.

This theory also has a local symmetry generated by the combined
gauge transformation
\begin{equation}
\delta_{\cal T}(\alpha)=\delta_Z(\omega)+\delta_X(\lambda)
\end{equation}
where the $\delta_Z(\omega)$ are the globally defined right action $G_R$ on the internal manifold ${\cal X}=G/ \G$ and the $\delta_X(\lambda)$
are antisymmetric tensor transformations acting on the $B$-field, where $\alpha_A=(\omega^m, \lambda_m)$ is the $O(d,d)$ covariant gauge
parameter of these transformations. In \cite{Hull:2005hk} the algebra of these infinitesimals was found to be
\begin{eqnarray}\label{Odd algebra}
\left[\delta_{\cal T}(\widetilde{\alpha}),\delta_{\cal
T}(\alpha)\right]
&=&\delta(t_{BC}{}^A\alpha^B\widetilde{\alpha}^C)
-\delta_W(L_{[A|D}t_{|BC]}{}^A\alpha^B\widetilde{\alpha}^C{\cal
A}^D)\nonumber\\
\left[\delta_W(\Lambda_{(1)}),\delta_{\cal T}(\alpha)\right] &=&0\nonumber\\
\left[\delta_W(\widetilde{\Lambda}_{(1)}),\delta_W(\alpha)\right]&=&0
\end{eqnarray}
where $\delta_W(\lambda_{(1)})$ generates antisymmetric tensor gauge transformations with the one-form parameter $\lambda_{(1)}$. The
characteristic field dependence in the commutator is a consequence of Chern-Simons terms of the form (\ref{Chern-simons field strength}) and
arises in a similar way to that seen in section 3. The Lie algebra subgroup of (\ref{Odd algebra}), analogous to that of section 3.2, is
\begin{eqnarray}\label{geometric algebraa}
\left[Z_m,Z_n\right]&=&-f_{mn}{}^pZ_p+K_{mnp}X^p\nonumber\\
\left[Z_m,X^n\right]&=&f_{mp}{}^nX^p\nonumber\\
\left[X^m,X^n\right]&=&0
\end{eqnarray}
where $Z_m$ generators of the right action $G_R$ on the twisted torus, as in section 3 and $X^m$ are generators of the antisymmetric tensor
transformations $B\rightarrow B+d\lambda$ and $m=1,2,3...d$. Combining the generators $Z_m, X^m$, where $m=1,2,3...d$, into an $O(d,d)$ vector
\begin{equation}
{\cal T}_A= \left(\begin{array}{cc} Z_m & X^m
\end{array}\right)
\end{equation}
the Lie algebra
may be
written as
\begin{equation}\label{Odd lie algebra}
\left[ {\cal T}_A,{\cal T}_B \right]=t_{AB}{}^C{\cal T}_C
\end{equation}

The gauge generators ${\cal T}_A$ are given in terms of the   $O(d,d)$ generators
$J^A{}_B$
by an expression of the form
\begin{equation}
{\cal T}_A=\Theta_{AB}{}^{C}J^B{}_C
\end{equation}
Here  $\Theta$ is
  the embedding tensor specifying the embedding of the gauge group into $O(d,d)$.
The generators can be used to define $J_{AB}=-J_{BA}= L_{AC}J^C{}_B$,
which
  satisfy the algebra
\begin{equation}
[J_{AB},J_{CD}]=L_{AD}J_{BC}+L_{BC}J_{AD}-L_{AC}J_{BD}-L_{BD}J_{AC},
\end{equation}
 In the case at hand, the embedding tensor can be read off explicitly.
 The generators $J^A{}_B$ decompose into the $GL(d,\R)$
 generators
 $J^a{}_b, J_a{}^b, J^{ab}, J_{ab}$
 and we find the gauge generators are
\begin{eqnarray}\label{embeds}
Z_m&=&f_{mn}^pJ^n{}_p-f_{mn}^pJ_p{}^n- \frac {1}{2}K_{mnp}J^{np}\nonumber\\
X^m&=& \frac {1}{2}f_{np}^mJ^{np}
\end{eqnarray}
and the embedding tensor can be read off from this.
It is completely specified by the choice of twist and flux.

\subsection{Heterotic Theory and $O(d,d+16)$}

The previous section is a truncation of the results
found in \cite{Odd} for the heterotic theory.
The low energy effective  Lagrangian  for the bosonic sector of the heterotic theory is
\begin{equation}
{\cal L}_{10}=e^{-\widehat{\Phi}}\left(\widehat{R}*1+d*\widehat{\Phi}\wedge\widehat{\Phi}
-\frac{1}{2}*\widehat{H}_{(3)}\wedge\widehat{H}_{(3)}-\frac{1}{2}tr\left(*\widehat{F}\wedge\widehat{F}\right)\right)
\end{equation}
where $a,b=1...16$ is a gauge index for the $E_8\times E_8$ or $Spin(32)/{\Z_2}$ gauge symmetry and $f_{ab}{}^c$ are the structure constants and
the trace is taken over the gauge indices. Setting $\alpha'=1$ the field strengths are
\begin{eqnarray}
\widehat{F}_{(2)}^a&=&d\widehat{A}_{(1)}^a+\frac{1}{2}f_{bc}{}^a\widehat{A}^b_{(1)}\wedge \widehat{A}^c_{(1)}\nonumber\\
\widehat{H}_{(3)}&=&d\widehat{B}_{(2)}-\frac{1}{2}tr\left(\widehat{A}_{(1)}\wedge
d\widehat{A}_{(1)}+\frac{2}{3}\widehat{A}_{(1)}\wedge\widehat{A}_{(1)}\wedge \widehat{A}_{(1)}\right)
\end{eqnarray}
The problem of adding flux to $\widehat{B}_{(2)}$ is greatly simplified by assuming the generators of the gauge group lie in the Cartan
subalgebra, breaking the gauge symmetry $E_8\times E_8$ or $Spin(32)/{\Z_2}\rightarrow U(1)^{16}$ for which $f_{ab}{}^c=0$. The reduction ansatz
for field strengths $\widehat{H}_{(3)}$ and $F_{(2)}^a$ are generalised to include the fluxes
\begin{eqnarray}
\widehat{H}_{(3)}&=&\frac{1}{6}K_{mnp}\sigma^m\wedge\sigma^n\wedge\sigma^p+...\nonumber\\
F^a_{(2)}&=&\frac{1}{2}M_{mn}^a\sigma^m\wedge\sigma^n+...
\end{eqnarray}
where $K_{mnp}$ and $M^a_{mn}$ are constant\footnote{There are some subtleties in adding a flux to $\widehat{B}_{(2)}$ in a way that preserves
the consistency of the Scherk-Schwarz truncation. See \cite{Odd} for details} and satisfy
\begin{equation}
M^a_{[mn|t}f_{|pq]}{}^t=0\qquad 2K_{[mn|t}f_{|pq]}{}^t=\delta_{ab}M^a_{[mn}M^b_{pq]}
\end{equation}
The reduced Lagrangian takes the same form as (\ref{O(d,d) Lagrangian}) except it is written in terms of $O(d,d+16)$ covariant fields. In
particular, the scalars parameterise the coset $O(d,d+16)/O(d)\times O(d+16)$
\begin{equation} {\cal M}^{AB}= \left(\begin{array}{ccc}
g^{mn} & -b_{(0)np}g^{pm} & -g^{mn}A_{(0)n}{}^a \\
-b_{(0)mp}g^{np} &
g_{mn}+g^{pq}b_{(0)mp}b_{(0)nq}+\delta_{ab}A_{(0)m}{}^aA_{(0)n}{}^b
& A_{(0)m}{}^a+b_{(0)mp}g^{pn}A_{(0)n}{}^a \\ -A_{(0)n}{}^ag^{mn}
& A_{(0)m}{}^a+A_{(0)n}{}^ag^{np}b_{(0)mp} &
\delta^{ab}+A_{(0)m}{}^ag^{mn}A_{(0)n}{}^b
\end{array}\right)
\end{equation}
where $b_{(0)mn}=B_{(0)mn}+\frac{1}{2}\delta_{ab}A_{(0)m}{}^aA_{(0)n}{}^b$ and $A=1,2,..2d+16$. There are also the $O(d,d+16)$ vector ${\cal
A}^A$ and corresponding field strength ${\cal F}^A$
\begin{equation}{\cal A}^A= \left(\begin{array}{ccc}
A^m \\ B_{(1)m} \\ A_{(1)}^a
\end{array}\right)  \qquad  {\cal F}^A= \left(\begin{array}{ccc}
F^m \\ G_{(2)m} - B_{(0)mn}F^n \\ F_{(2)}^a
\end{array}\right)
\end{equation}
and the $O(d,d)$ invariant matrix $L_{AB}$ is replaced by the
$O(d,d+16)$ invariant
\begin{equation} L_{AB}=\left(\begin{array}{ccc}
0 & \bid_d & 0 \\ \bid_d & 0 & 0 \\ 0 & 0 & \bid_{16}
\end{array}\right) \end{equation}

As in the previous section, the gauging breaks the global symmetry, but the
Lagrangian is formally invariant under the action of the global $O(d,d+16)$ if the structure constants of the gauge
group transform as $O(d,d+16)$ tensors. The $O(d,d+16)$ rigid symmetry then maps
one gauging into another, in which the gauge algebra remnains the same, but its is embedding
in the duality group changes.

The gauge group $L\subset O(d,d+16)$ has symmetry algebra
\begin{eqnarray}\label{O(d,d+16) algebra}
\left[\delta_Z(\widetilde{\omega}),\delta_Z(\omega)\right]&=&\delta_Z(f_{mn}{}^p\omega^m\widetilde{\omega}^n)
-\delta_Y(\delta_{ab}M^b_{mn}\omega^m\widetilde{\omega}^n) -\delta_X(K_{mnp}\omega^n\widetilde{\omega}^p)
-\delta_X(K_{mnp}\omega^n\widetilde{\omega}^pA^m)\nonumber\\
\left[\delta_Y(\lambda^a_{(0)}),\delta_Z(\omega)\right]&=& -\delta_X(\delta_{ab}M^b_{mn}\lambda^a_{(0)}\omega^n)
-\delta_X(\delta_{ab}M^b_{mn}\lambda^a_{(0)}\omega^nA^m)\nonumber\\
\left[\delta_X(\lambda_{(0)}m),\delta_Z(\omega^n)\right]&=&
\delta_X(f_{mn}{}^p\lambda_{(0)_p}\omega^n)
\end{eqnarray}
where $\delta_Y(\lambda^a_{(0)})=\lambda^a_{(0)}Y_a$ generates the infinitesimal gauge transformation $\delta A^a_{(1)}=d\lambda^a_{(0)}$. All
other commutators are zero. The symmetry algebra (\ref{c-field algebra}) contains the Lie subalgebra first identified in \cite{Odd}
\begin{eqnarray}\label{O(d,d+16) Lie algebra}
\left[Z_m,Z_n\right]&=&-f_{mn}{}^pZ_p-M^a_{mn}Y_a+K_{mnp}X^p\nonumber\\
\left[X^m,Z_n\right]&=&-f_{np}{}^mX^p\nonumber\\
\left[Y_a,Z_m\right]&=&-\delta_{ab}M^b_{mn}X^n\nonumber\\
\left[Y_a,Y_b\right]&=&\left[Y_a,X^m\right]=\left[X^m,X^n\right]=0
\end{eqnarray}
This algebra may be written in an $O(d,d+16)$ covariant form
(\ref{Odd lie algebra}) where the generators form an $O(d,d+16)$
vector
\begin{equation}
{\cal T}_A= \left(\begin{array}{ccc} Z_m & X^m & Y_a
\end{array}\right)
\end{equation}
The symmetry algebra (\ref{O(d,d+16) algebra}) and Lie subalgebra (\ref{O(d,d+16) Lie algebra}) can then be written in the $O(d,d+16)$ covariant
form of (\ref{Odd algebra}) and (\ref{Odd lie algebra}) respectively, where the structure constants $t_{AB}{}^C$ are given by
$t_{mn}{}^p=f_{mn}{}^p$, $t_{mn}{}^a=M_{mn}{}^a$ and $t_{mnp}=K_{mnp}$.

\subsection{M-Theory, Gauged Supergravity and $E_{d(d)}$}

Dimensional reduction of a generic field theory coupled to
gravity on $\mathcal{X}\simeq T^d$ followed by Kaluza-Klein truncation to the zero modes  has a global $GL(d,\R)$ symmetry. Reductions of eleven dimensional supergravity have at least a global $GL(d,\R)\ltimes \R^q$ where the
$\R^q$ comes from constant shifts of the three form potential and $q=\frac{1}{6}d(d-1)(d-2)$. The reductions considered in section three gauge a
subgroup of $GL(d,\R)\ltimes \R^q$. Dualising all $p$-form gauge
fields with degree $p>\frac{1}{2}D$ in the ungauged theory changes the global symmetry to a global $E_{d(d)}$ symmetry\footnote{For
$d=6,7,8$, $E_{d(d)}$ are the exceptional groups $E_{6(6)}$, $E_{7(7)}$ and $E_{8(8)}$ and for $d=2,3,4,5$ the $E_{d(d)}$ groups are defined as
$SL(2;\R)\times O(1,1)$, $SL(3;\R)\times SL(2;\R)$, $SL(5;\R)$ and $O(5,5)$ respectively.} \cite{Cremmer:1979up,Julia:1980gr}.
 For reductions to odd dimensions, $E_{d(d)}$ is a symmetry of the Lagrangian.
 For reductions to  even
dimensions, $E_{d(d)}$ is a symmetry
of the equations of motion and Bianchi identities
\cite{Cremmer:1997ct,Cremmer:1998px,Hull:2003kr,deWit:2005ub}. However one may introduce an auxiliary Lagrangian, using the doubled formalism of
\cite{Cremmer:1997ct,Cremmer:1998px}, in which the field strength of degree $\frac{1}{2}D$ is combined with its dual into a single irreducible
representation of $E_{d(d)}$. The number of physical degrees of freedom is maintained by requiring that this doubled field satisfy a twisted
self-duality condition \cite{Cremmer:1997ct,Hull:2003kr}. Using these doubled fields, a Lagrangian for this theory may be constructed with
manifest $E_{d(d)}$ global symmetry.

Given a supergravity with an $E_{d(d)}$ global symmetry of the action (which uses the doubled formalism in even dimensions),
 one may then seek to supersymmetric gaugings of  subgroups of $E_{d(d)}$. Many classes of examples have been found, but until recently no coherent framework for
a programme to systematically classify such gauged supergravities had been found. In
\cite{deWit:2004nw,deWit:2005hv,Samtleben:2005bp,deWit:2005ub} doubled Lagrangians were proposed for the $D=5$ and $D=7$ gauged supergravitites
with manifest $E_{6(6)}$ and $SL(5)$ covariance respectively. Similar actions are conjectured to exist in all dimensions. These `universal'
Lagrangians are conjectured to contain all possible gaugings of $E_{d(d)}$ and are reviewed in the following sections.
The gauged supergravities discussed here that arise from twisted torus reductions with flux arise from
gauging subgroups of the $GL(d,\R)\ltimes \R^q$ that is a symmetry before dualising.
Moreover, the non-abelian interactions of the gauged supergravity provide obstructions to the    dualisations of $p$-form gauge fields used in the ungauged theory to obtain the $E_{d(d)}$ symmetric form.
However, instead of dualising $p$-form gauge fields  to $D-p-2$ form gauge fields one can instead use the doubled form with both  $p$-form gauge fields and $D-p-2$ form gauge fields, using a \lq universal' lagrangian for the gauged supergravity.
This raises the question as to whether the  gaugings obtained from twisted torus reductions can fit into the class of gaugings of subgroups of $E_{d(d)}$ of \cite{deWit:2004nw,deWit:2005hv,Samtleben:2005bp,deWit:2005ub}, or whether they provide a separate universality class.
At first glance, the lagrangians seem to be of a rather different form, with our compactifications giving second order kinetic terms for certain $p$-form gauge fields, whereas the corresponding  $p$-form gauge fields
in the  $E_{d(d)}$-covariant formulation  have a
 first order kinetic term. We show that these kinetic terms are in fact dual actions for the same theory.
 The theories then agree at the quadratic level and have the same supersymmetry and gauge symmetry, so the full non-linear theories should be identical. We check this explicitly in a particular case, and
 provide a number of    checks on the conjecture that the full theories arising from twisted torus compactifications with flux do indeed arise as gaugings of    $E_{d(d)}$ in the universal approach.

\subsubsection{ Five Dimensional Gauged Supergravity }

Compactifying eleven dimensional supergravity on a six dimensional torus and dualising the $C_{(3)}$ and $C_{(2)m}$ potentials gives a theory
with a rigid $E_{6(6)}$ symmetry and one form potentials ${\cal A}^A_{(1)}$ ($A=1,....,27$) transforming in the $\mathbf{\overline{27}}$ of $E_{6(6)}$. A
gauging of the theory, which breaks the global $E_{6(6)}$ symmetry, sees some of the ${\cal A}^A_{(1)}$ become non-abelian gauge bosons, whilst
those that are not involved in the gauging must be dualised to massive, self-dual two forms ${\cal B}_{(2)A}$ \cite{Gunaydin:1985cu}. The difficulty
in performing a systematic analysis of the different gaugings is due, in part, to the fact that different gaugings require different numbers of
one forms to be dualised and therefore the Lagrangians of differing gaugings may appear quite different. This problem was overcome in
\cite{deWit:2004nw} where a doubled formalism was proposed in which 27 ${\cal B}_{(2)A}$ are introduced in addition to the 27 ${\cal A}^A_{(1)}$
and general gauge groups are allowed. The correct number of physical degrees of freedom is maintained by introducing additional gauge symmetries
which remove the extra unphysical degrees of freedom. When a choice of gauge group is made, the excess ${\cal B}_{(2)A}$ are projected out and
the ${\cal A}^A_{(1)}$ not involved in the gauging are eaten by the remaining ${\cal B}_{(2)A}$. This five dimensional example, reviewed below,
was studied at length in \cite{deWit:2004nw} where further details may be found.

The subgroup of the global $E_{6(6)}$ symmetry that is to be gauged is specified by
an embedding tensor $\Theta_A{}^{\alpha}$, giving the   generators ${\cal T}_A$ of the gauge group $G$ in terms of the global symmetry
generators $J_{\alpha}$ of $E_{6(6)}$
\begin{equation}\label{embedding tensor}
{\cal T}_A=\Theta_A{}^{\alpha}J_{\alpha}
\end{equation}
The gauge algebra is required to close to give
\begin{equation}
[{\cal T}_A,{\cal T}_B]=t_{AB}{}^C{\cal T}_C
\end{equation}
for some $t_{AB}{}^C$. Consistency of the gauging and the requirement of maximal supersymmetry place constraints on which gaugings and groups
are allowed and these constraints were studied in detail in \cite{deWit:2005ub,deWit:2005hv,Samtleben:2005bp,deWit:2004nw}. One of these is that the embedding tensor
is required to be in the $\mathbf{351}$ representation of $E_{6(6)}\times E_{6(6)}$.
 The ungauged theory has one-form fields ${\cal
A}_{(1)}{}^A$ which transform in the $\mathbf{\overline{27}}$ of $E_{6(6)}$, and under the action of the gauge symmetry with parameters
$\Lambda_{(0)}{}^A(x)$, these one-forms transform as connections, up to terms annihilated by  projection with the embedding tensor
\begin{equation}\label{hello}
\Theta_A{}^{\alpha}\left(\delta_{\Lambda}(\Lambda_{(0)}){\cal
A}_{(1)}{}^A\right)=\Theta_A{}^{\alpha}\left(D\Lambda_{(0)}{}^A\right)
\end{equation}
where the derivative is given by
\begin{equation}
D\Lambda^A_{(0)}=d{\cal A}^A_{(1)}+g{\cal T}_{BC}{}^A\Lambda^B_{(0)}{\cal A}^C_{(1)}
\end{equation}
 It is useful to define the following matrix representation of the
gauge group generators acting on the 27-dimensional representation  $({\cal T}_A)_B{}^C\equiv {\cal T}_{AB}{}^C$ where it is stressed that ${\cal T}_{AB}{}^C$ will not be antisymmetric in
the lower indices in general. For the gauging defined by the embedding (\ref{embedding tensor}) to be consistent $({\cal T}_A)_B{}^C$ must
decompose into the adjoint representation of the gauge group plus parts that vanish under contraction with the embedding tensor such that
\begin{equation}\label{projection}
({\cal
T}_A)_B{}^C\Theta_C{}^{\alpha}=-t_{AB}{}^C\Theta_C{}^{\alpha}
\end{equation}
Using the totally symmetric $E_{6(6)}$ invariant tensors $d_{ABC}$ and $d^{ABC}$, the tensor ${\cal Z}^{AB}=-{\cal Z}^{BA}$ is defined as
\begin{equation}
{\cal Z}^{AB}={\cal T}_{CD}{}^{[A}d^{B]CD}
\end{equation}
such that
\begin{equation}
{\cal T}_{(AB)}{}^C=d_{ABD}{\cal Z}^{CD}
\end{equation}
Furthermore it may be shown that ${\cal Z}^{AB}\Theta_B{}^{\alpha}={\cal Z}^{AB}{\cal T}_B=0$. These constraints then ensure that ${\cal
T}_{(AB)}{}^C\Theta_C{}^{\alpha}=d_{ABD}{\cal Z}^{DC}\Theta_C{}^{\alpha}=0$ so that the ${\cal T}_{AB}{}^C\Theta_C{}^{\alpha}$ are
antisymmetric  ${\cal T}_{AB}{}^C\Theta_C{}^{\alpha}=-{\cal T}_{BA}{}^C\Theta_C{}^{\alpha}$, and they are to are identified with the structure constants of the gauge group.

As explained in \cite{deWit:2004nw}, the required generalisation of the gauge transformation is
\begin{equation}
\delta_{\Lambda}(\Lambda_{(0)}){\cal A}_{(1)}{}^A=d\Lambda_{(0)}{}^A-g{\cal T}_{[BC]}{}^A\Lambda_{(0)}{}^C{\cal A}_{(1)}{}^B-gZ^{AB}\Xi_{(1)B}
\end{equation}
where a shift symmetry with arbitrary parameter $\Xi_{(1)A}(x)$ has been introduced. This indeed projects to (\ref{hello}).
 The gauge fixing of
this symmetry ensures that the number of degrees of freedom in this doubled formalism reduces to the correct number with the shift symmetry
removing the surplus degrees of freedom. The following results are also useful
\begin{eqnarray}
{\cal T}_{AC}{}^D{\cal T}_{BD}{}^E-{\cal T}_{BC}{}^D{\cal T}_{AD}{}^E+{\cal T}_{AB}{}^D{\cal T}_{DC}{}^E&=&0\nonumber\\
{\cal T}_{[AB]}{}^C{\cal T}_{[DC]}{}^E+{\cal T}_{[DA]}{}^C{\cal
T}_{[BC]}{}^E+{\cal T}_{[BD]}{}^C{\cal
T}_{[AC]}{}^E&=&d_{FC[D}{\cal T}_{AB]}{}^C{\cal Z}^{FE}
\end{eqnarray}
so that the $T_{[AB]}{}^C$ only satisfy the Jacobi identity in the subspace projected by the embedding tensor.

\subsubsection{$E_{6(6)}$ Covariant Lagrangian}

The bosonic sector of the $E_{6(6)}$ universal Lagrangian is, to quadratic order,
\begin{eqnarray}\label{Universal Lagrangian}
{\cal L}_5&=&R*1+\frac{1}{4}tr\left(*D{\cal M}\wedge D{\cal M}^{-1}\right)
\nonumber\\
&&-\frac{1}{2}{\cal M}_{AB}*{\cal H}_{(2)}{}^A\wedge {\cal H}_{(2)}{}^B+\frac{1}{2}g{\cal Z}^{AB}{\cal B}_{(2)A}\wedge D{\cal B}_{(2)B}+...
\end{eqnarray}
where $+...$ denotes terms of higher order. The ${\cal M}$ are scalars parameterising the coset space $E_{6(6)}/USp(8)$. The two-form field
strength is
\begin{equation}
{\cal H}_{(2)}{}^A=d{\cal A}_{(1)}{}^A-\frac{1}{2}g{\cal
T}_{[BC]}{}^A{\cal A}_{(1)}{}^B\wedge {\cal A}_{(1)}{}^C+g{\cal
Z}^{AB}{\cal B}_{(2)N}
\end{equation}
This field strength transforms covariantly as
\begin{equation}
\delta (\Lambda_{(0)}){\cal H}_{(2)}{}^A=-g{\cal
T}_{CB}{}^A\Lambda_{(0)}{}^B{\cal H}_{(2)}{}^C
\end{equation}
under the gauge symmetry generated by the infinitesimal
transformations
\begin{eqnarray}
{\cal Z}^{AB}\delta {\cal B}_{(2)B}&=&{\cal Z}^{AB}D\Xi_{(1)B}+g{\cal Z}^{AB}\Lambda_{(0)}{}^C{\cal T}_{[CB]}{}^D{\cal B}_{(2)D}
\nonumber\\
&&-g{\cal Z}^{AB}\left(d_{BCD}d{\cal A}_{(1)}{}^C-\frac{1}{2}g{\cal T}_{EB}{}^Cd_{CDF}{\cal A}_{(1)}{}^E\wedge {\cal A}_{(1)}{}^F\right)
\nonumber\\
\delta {\cal A}_{(1)}{}^A&=&d\Lambda_{(0)}{}^A-g{\cal T}_{[BC]}{}^A\Lambda_{(0)}{}^B{\cal A}_{(1)}^C-g{\cal Z}^{AB}\Xi_{(1)B}
\end{eqnarray}
These infinitesimals generate the symmetry algebra
\begin{eqnarray}
\left[\delta_{\Lambda}(\widetilde{\Lambda}_{(0)}),\delta_{\Lambda}(\Lambda_{(0)})\right]&=&\delta_{\Lambda}\left(g{\cal
T}_{[BC]}{}^A\widetilde{\Lambda}_{(0)}{}^B\Lambda_{(0)}{}^C\right)-\delta_{\Xi}\left(gd_{AB[C}{\cal
T}_{DE]}{}^B\widetilde{\Lambda}_{(0)}{}^C\Lambda_{(0)}^D{\cal
A}_{(1)}{}^E\right)
\nonumber\\
\left[\delta_{\Xi}(\Xi_{(1)}),\delta_{\Lambda}(\Lambda_{(0)})\right]&=&\delta_{\Xi}\left(\frac{1}{2}g{\cal
T}_{CM}{}^B\Lambda_{(0)}{}^C\Xi_{(1)B}\right)
\nonumber\\
\left[\delta_{\Xi}(\widetilde{\Xi}_{(0)}),\delta_{\Xi}(\Xi_{(0)})\right]&=&0
\end{eqnarray}
This symmetry algebra is not a Lie algebra due to the field dependence on the right hand side of the first commutator and is of the general form
of the algebras found in flux compactifications of field theories on twisted tori (\ref{c-field algebra}).

The Lagrangian (\ref{Universal Lagrangian}) is conjectured to describe all possible gaugings of maximal supergravity in five dimensions, where a
specific gauged supergravity is defined by the appropriate choice of embedding tensor $\Theta_A{}^{\alpha}$. The fields in (\ref{Universal
Lagrangian}) transform covariantly under the action of the global $E_{6(6)}$, but for a given gauging the Lagrangian is not invariant. However,
if we allow the embedding tensor, and in particular ${\cal T}_{[AB]}{}^C$ and ${\cal Z}^{AB}$, to transform under $E_{6(6)}$ then the Lagrangian
(\ref{Universal Lagrangian}) \emph{is} invariant. The action of the global $E_{6(6)}$ changes the   embedding tensor, relating
apparently different gauged supergravities to each other.
One may then think of the $E_{6(6)}$-invariant Universal Lagrangian (\ref{Universal Lagrangian}) for gauged supergravity as the counterpart of the $O(d,d)$ and
$O(d,d+16)$ theories (\ref{O(d,d) Lagrangian}).

The gauged supergravities in seven dimensions have a similar structure, with an $SL(5,\R)$ invariant action \cite{Samtleben:2005bp}. For  even
dimensions, the self-duality of field strengths of degree $D/2$ makes the construction of the $O(5,5)$ and $E_{7(7)}$ invariant Lagrangians with
doubled degrees of freedom in six and four dimensions more challenging, but recent progress \cite{deWit:2005hv,deWit:2005ub} suggests that the
results of \cite{deWit:2004nw} and \cite{Samtleben:2005bp} can be extended to all dimensions.

\subsubsection{Symmetry Breaking and Gauged Supergravity}

The choice of an embedding tensor breaks the $E_{6(6)}$ invariance of the universal Lagrangian (\ref{Universal Lagrangian}). Following
\cite{deWit:2004nw}, an $E_{6(6)}$ basis may be chosen where $A=(m,a,u)$ with $m=1,2,..s$, $a=s+1,s+2,..27-t$ and $u=28-t,29-t,..27$ where $s$
is the rank of the embedding tensor and $t$ is the rank of ${\cal Z}^{AB}$. In this basis the constant $E_{6(6)}$ tensors are written
\begin{equation}
({\cal T}_m)_B{}^C=\left(%
\begin{array}{ccc}
  -t_{mn}{}^p & h_{mn}{}^a & C_{mn}{}^u \\
  0 & 0 & C_{ma}{}^u \\
  0 & 0 & D_{mv}{}^u \\
\end{array}%
\right)\qquad   {\cal Z}^{AB}=\left(%
\begin{array}{ccc}
  0 & 0 & 0 \\
  0 & 0 & 0 \\
  0 & 0 & {\cal Z}^{uv} \\
\end{array}%
\right)
\end{equation}
where ${\cal Z}^{uv}$ is non-degenerate and invertible.
This defines the gauge algebra
\begin{eqnarray}\label{algy7}
\left[T_m,T_n\right]&=&-t_{mn}{}^pT_p+h_{mn}{}^aT_a+C_{mn}{}^uT_u\nonumber\\
\left[T_m,T_a\right]&=&C_{ma}{}^uT_u\nonumber\\
\left[T_m,T_u\right]&=&D_{mu}{}^vT_v
\end{eqnarray}
which is indeed a subalgebra of $E_{6(6)}$.

All antisymmetric tensors ${\cal B}_{(2)A}$ appear in the Lagrangian contracted with
${\cal Z}^{AB}$, so the above choice of coordinates project out all but the ${\cal B}_{(2)u}$ from the theory. Making the gauge choice
\begin{equation}
\Xi_{(1)u}=g^{-1}{\cal Z}_{uv}{\cal A}_{(1)}{}^v
\end{equation}
the gauge bosons ${\cal A}_{(1)}{}^u$ are gauged to zero, whilst the ${\cal B}_{(2)u}$ absorb the ${\cal A}_{(1)}{}^u$ degrees of freedom.
Defining the tensor
\begin{equation}
\breve{{\cal B}}_{(2)}{}^u={\cal Z}^{uv}{\cal
B}_{(2)v}+g^{-1}D{\cal A}_{(1)}{}^u+...
\end{equation}
the gauged theory, to quadratic order, becomes
\begin{eqnarray}\label{gauge fixed lagrangian}
{\cal L}_5&=&R*1+\frac{1}{4}tr\left(*D{\cal M}\wedge D{\cal M}^{-1}\right)
\nonumber\\
&&-\frac{1}{2}{\cal M}_{AB}*{\cal F}_{(2)}{}^A\wedge {\cal F}_{(2)}{}^B+\frac{1}{2}g{\cal Z}_{uv}\breve{{\cal B}}_{(2)}{}^u\wedge D\breve{{\cal
B}}_{(2)}{}^v+...
\end{eqnarray}
where
\begin{equation}
{\cal F}_{(2)}{}^A=\left(%
\begin{array}{c}{\cal F}_{(2)}{}^m \\ {\cal F}_{(2)}{}^a \\ \breve{{\cal B}}_{(2)}{}^u \\
\end{array}%
\right)
\end{equation}
and ${\cal F}_{(2)}$ are covariant field strengths.
\begin{eqnarray}
{\cal F}_{(2)}{}^m&=&d{\cal A}^m-\frac{1}{2}gt_{np}{}^m{\cal
A}_{(1)}{}^n\wedge {\cal A}_{(1)}{}^p\nonumber\\
{\cal F}_{(2)}{}^a&=&d{\cal A}^a+\frac{1}{2}gh_{mn}{}^a{\cal A}_{(1)}{}^m\wedge {\cal A}_{(1)}{}^n
\end{eqnarray}
The Lagrangian (\ref{gauge fixed lagrangian}) has a gauge symmetry with  Lie algebra
\begin{equation}
\left[T_m,T_n\right]=-t_{mn}{}^pT_p+h_{mn}{}^aT_a
\end{equation}
where all other commutators vanish.
Note that this is not a subalgebra of  (\ref{algy7}) in general, but is the remaining symmetry after the gauge fields ${\cal A}_{(1)}{}^u$ have been eliminated.

\subsection{Flux Compactifications of Eleven Dimensional Supergravity and the Universal Lagrangian}

Setting the flux and geometric twists to zero, the reductions of section 2 give the reduction of eleven-dimensional supergravity on $T^d$. The
resulting effective theory is a massless, ungauged, maximal supergravity in $D$-dimensions with $GL(d;\R)\ltimes \R^{d(d-1)(d-2)/6}$ global
symmetry. Dualisation takes the ungauged, massless theory that arises from dimensional reduction to a theory with a global $E_{d(d)}$ symmetry.
For more general reductions, such as those considered in this paper, fluxes and curvature of the internal space give rise to massive
deformations and in general
one finds obstructions to the usual dualisation procedure. 
There is then an issue of whether the Lagrangians produced by flux compactifications on twisted tori presented here arise within the
Universal Lagrangian formalism. We shall argue that the Lagrangian (\ref{reduced Lagrangian}) is not described by the Universal Lagrangians, but an equivalent, dual
form can be found which \emph{is} contained in the Universal formalism.
First, we shall discuss the dualisations needed for the discussion.

\subsubsection{Dualisation}

It is instructive to begin by reviewing the Hodge dualisation \cite{Giveon:1994mw}. Consider a $D$-dimensional  $(p-1)$-form gauge theory where the
field strength $G_{(p)}=dC_{(p-1)}$ is given in terms of the potential $C_{(p-1)}$. The equations of motion and Bianchi identity
\begin{equation}
d*G_{(p)}=0 \qquad  dG_{(p)}=0
\end{equation}
are exchanged under the duality generated by $G_{(p)}\rightarrow
H_{(D-p)}=*G_{(p)}$. The dual theory then has equations of motion
and Bianchi identity
\begin{equation}
d*H_{(D-p)}=0 \qquad  dH_{(D-p)}=0
\end{equation}
The dual Bianchi identity allows one to define a potential $\vartheta_{(D-p-1)}$ locally  such that $H_{(D-p)}=d\vartheta_{(D-p-1)}$. This
duality may be derived from a Lagrangian by treating the field strength $G_{(p)}$ as the independent variable and introducing
$\vartheta_{(D-p-1)}$ as a Lagrange multiplier, constraining the theory to satisfy the Bianchi identity $dG_{(p)}=0$. The Lagrangian is
\begin{equation}\label{Lagrangian 7}
{\cal L}=-\frac{1}{2}*G_{(p)}\wedge
G_{(p)}+d\vartheta_{(D-p-1)}\wedge G_{(p)}
\end{equation}
Variation with respect to $G_{(p)}$ leads to the duality constraint $*G_{(p)}=H_{(D-p)}$. Substituting for $G_{(p)}$ back into the Lagrangian
gives the Lagrangian for the dual theory.\footnote{In fact, the dual Lagrangian calculated in this way is ${\cal
L}=-\frac{1}{2}(-1)^{p(D-p)}*H_{(D-p)}\wedge H_{(D-p)}$ We shall assume that $\vartheta_{(D-p-1)}$ is rescaled to
$\xi\vartheta_{(D-p-1)}$ where $\xi^2(-1)^{p(D-p)}=-1$ to give the kinetic term the canonical normalisation.}

More general models, particularly those based on reductions of
eleven-dimensional supergravity, will have  Chern-Simons
terms in the field strengths and in the
Lagrangian. It is therefore necessary to generalise the above toy
model to include such terms. Consider a field strength
\begin{equation}\label{G-field}
G_{(p)}=dC_{(p-1)}+{\cal W}_{(p)}
\end{equation}
with a Chern-Simons like term ${\cal W}_{(p)}$   that is not closed in general and  the $D$-dimensional Lagrangian
\begin{equation}\label{Lagrangian 1}
{\cal L}=-\frac{1}{2}*G_{(p)}\wedge G_{(p)}+{\cal Q}_{(q)}\wedge
dC_{(p-1)}+...
\end{equation}
where ${\cal Q}_{(q)}$ and terms denoted by $+...$ are independent of $C_{(p-1)}$. Following \cite{Lu:1995yn}, we will refer to  $q$-form ${\cal Q}_{(q)}$ with $q=D-p$
as a transgression term; in general, it is not closed. The field strength satisfies the Bianchi identity
\begin{equation}\label{Bianchi}
dG_{(p)}=d{\cal W}_{(p)}
\end{equation}
and the Chern-Simons term has the property that $d{\cal W}_{(p)}$ transforms covariantly under the gauge group even though ${\cal W}_{(p)}$ generally will not. The reduction of the
eleven-dimensional supergravity on a torus, followed by a truncation to the zero modes, gives a Lagrangian of this form.

 In
the toy model above with ${\cal W}={\cal Q}=0$,   duality exchanges the equations of motion and Bianchi identity. The exchange
$G_{(p)}\leftrightarrow *G_{(p)}$ is then a symmetry of the theory. In this more general case, it is no longer the case that $G_{(p)}$ is simply
exchanged with its Hodge dual. The correct duality transformation requires that ${\cal Q}$ and the Chern-Simons term ${\cal W}$ should also be
exchanged. Treating the field strength $G_{(p)}$ as an independent variable and adding a Lagrange multiplier term generalises (\ref{Lagrangian
7}) to the Lagrangian
\begin{equation}\label{Lagrangian 2}
{\cal L}_{G,\vartheta}=-\frac{1}{2}*G_{(p)}\wedge G_{(p)}+{\cal
Q}_{(q)}\wedge \left(G_{(p)}-{\cal
W}_{(p)}\right)+\vartheta_{(q-1)}\wedge d\left(G_{(p)}-{\cal
W}_{(p)}\right)+...
\end{equation}
Variation of this Lagrangian with respect to $\vartheta_{(q-1)}$ gives the Bianchi identity (\ref{Bianchi}), from which we may introduce the
$C_{(p-1)}$ potential as in (\ref{G-field}). This definition of the field strength (\ref{G-field}) may then be substituted back into
(\ref{Lagrangian 2}) and the equation of motion
\begin{equation}
d*G_{(p)}=d{\cal Q}_{(q)}
\end{equation}
arises from a subsequent variation with respect to $C_{(p-1)}$. If instead $G_{(p)}$ is treated as the independent variable, the variation of
the Lagrangian (\ref{Lagrangian 2}) with respect to $G_{(p)}$ is
\begin{equation}\label{constraint 3}
G_{(p)}=*H_{(q)}
\end{equation}
where the dual field strength is defined as
\begin{equation}
H_{(q)}=d\vartheta_{(q-1)}+{\cal Q}_{(q)}
\end{equation}
Substituting for $G_{(p)}$ using (\ref{constraint 3}) in the
Lagrangian (\ref{Lagrangian 2}) gives the dual formulation of the
theory
\begin{equation}\label{Lagrangian 3}
\widetilde{{\cal L}}=\frac{1}{2}*H_{(p)}\wedge H_{(p)}-{\cal
W}_{(p)}\wedge H_{(q)}
\end{equation}
The interchange of $\cal Q$ and ${\cal W}$ terms is quite clear from this example. This is a general feature of such dualisations.

This method
of dualisation can not be applied to the flux reductions of eleven-dimensional supergravity on $T^d$ as will now be demonstrated. In eleven
dimensions this theory has Chern-Simons term
\begin{equation}\label{CS term}
{\cal L}_{11}=\frac{1}{6}\widehat{G}\wedge \widehat{G}\wedge\widehat{C}
\end{equation}
If the three-form has a constant left-invariant flux of the form (\ref{flux}) then $\widehat{C}=C+\varpi_{(3)}$, where ${\cal K}=d\varpi_{(3)}$,
as in section 2. The Chern-Simons term (\ref{CS term}), after integrations by parts, becomes
\begin{equation}\label{flux CS}
{\cal L}_{11}=\frac{1}{6}dC\wedge dC\wedge C+\frac{1}{2}dC\wedge C\wedge {\cal K}+\frac{1}{2}C\wedge {\cal K}\wedge {\cal K} +\frac{1}{6}{\cal
K}\wedge {\cal K}\wedge\varpi_{(3)}
\end{equation}
The last term may be ignored here as it does not contribute to the equations of motion of $C$. Reducing (\ref{flux CS}) on $T^d$
one finds that the reduced theory includes terms of the form\footnote{See Appendix C for details.}
\begin{equation}\label{reduced CS term}
{\cal L}_D=\frac{1}{2}\mu C_{(p-1)}\wedge G_{(q+1)}+...
\end{equation}
where $D=p+q$ and $\mu$ is a constant parameter related to the constant flux\footnote{All internal frame indices have been suppressed but in
general there will be a contraction of these indices with an alternating symbol,  proportional to
\begin{equation}
\frac{1}{2}\mu^{m_1m_2...m_{4-p}n_1n_2...n_{3-q}}
C_{(p-1)m_1m_2...m_{4-p}}\wedge G_{(q+1)n_1n_2...n_{3-q}}
\end{equation}} $K_{mnpq}$. For example, in $D=7$
\begin{equation}
\mu=\frac{1}{24}\epsilon^{mnpq}K_{mnpq}
\end{equation}
Terms of the form (\ref{reduced CS term}) are not included in the Lagrangian (\ref{Lagrangian 2}) so the previous considerations must be
generalised in the presence of flux. Such terms are mass terms and occur in two distinct ways. The first type of mass term occurs when $p\neq
q+1$ and in principle may occur in any dimension. The second, where $p=q+1$, is the special case of a topologically massive theory and only
occur in odd dimensions. The following sections give explicit constructions of the dual formulations of such Lagrangians.

\subsubsection{Duality of Massive Theories}

Consider the Lagrangian
\begin{equation}\label{Lagrangian 4}
{\cal L}=-\frac{1}{2}*G_{(p)}\wedge
G_{(p)}-\frac{1}{2}*F_{(q+1)}\wedge F_{(q+1)}+\mu C_{(p-1)}\wedge
F_{(q+1)}
\end{equation}
for the potentials $C_{(p-1)}$ and $B_{(q-1)}$ with the
field strengths $G_{(p)}=dC_{(p-1)}$ and $F_{(q+1)}=dB_{(q)}$
where $D=p+q$ and $\mu$ is a constant. In terms of
(\ref{Lagrangian 1}) this Lagrangian has ${\cal Q}_{(q)}=\mu
B_{(q)}$. $F_{(q+1)}$ is dualised by introducing a dual potential
$\vartheta_{(p-2)}$ as a Lagrange multiplier, to enforce the
Bianchi identity $dF_{(q+1)}=0$. The constrained Lagrangian is
\begin{equation}\label{Lagrangian 6}
{\cal L}=-\frac{1}{2}*G_{(p)}\wedge
G_{(p)}-\frac{1}{2}*F_{(q+1)}\wedge F_{(q+1)}+\mu C_{(p-1)}\wedge
F_{(q+1)}-\vartheta_{(p-2)}\wedge dF_{(q+1)}
\end{equation}
Considering $F_{(p)}$ as the independent variable and varying
(\ref{Lagrangian 6}) with respect to it defines the dual field
strength $H_{(p-1)}=*F_{(q+1)}$ where
\begin{equation}
H_{(p-1)}=d\vartheta_{(p-2)}-\mu C_{(p-1)}
\end{equation}
The dual Lagrangian is
\begin{equation}
\widetilde{{\cal L}}=-\frac{1}{2}*G_{(p)}\wedge
G_{(p)}-\frac{1}{2}*H_{(p-1)}\wedge H_{(p-1)}
\end{equation}
This dual theory is invariant under the abelian gauge symmetry
generated by the infinitesimal variations
\begin{equation}
\delta C_{(p-1)}=d\lambda_{(p-2)}\qquad
\delta\vartheta_{(p-2)}=-\mu\lambda_{(p-2)}
\end{equation}
A massive gauge singlet potential may be defined as
\begin{equation}
S_{(p-1)}=C_{(p-1)}-\mu^{-1}d\vartheta_{(p-2)}
\end{equation}
such that $H_{(p-1)}=-\mu S_{(p-1)}$. The dual Lagrangian may then
be written
\begin{equation}\label{Lagrangian 6}
\widetilde{{\cal L}}=-\frac{1}{2}*dS_{(p-1)}\wedge
dS_{(p-1)}-\frac{1}{2}\mu^2*S_{(p-1)}\wedge S_{(p-1)}
\end{equation}
The reduction of the eleven-dimensional supergravity to six dimensions with flux contains terms of the form   (\ref{Lagrangian 4}) with $p=q=3$,
which may then be rewritten in the form (\ref{Lagrangian 6}). We anticipate that it is this latter form  that arises in the $O(5,5)$ covariant
Universal Lagrangian.

\subsubsection{Duality of Topologically Massive Theories}

Topologically massive theories  are possible in odd dimensions   for forms of degree $\frac{1}{2}(D-1)$\cite{Townsend:1983xs,Deser:1982vy,Deser:1981wh}. Consider
  the Lagrangian
\begin{equation}\label{Lagrangian 5}
{\cal L}_C=-\frac{1}{2}*G_{(p)}\wedge G_{(p)}+\frac{1}{2}\mu C_{(p-1)}\wedge dC_{(p-1)}
\end{equation}
where $G_{(p)}=dC_{(p-1)}$.    Flux reductions of
eleven-dimensional supergravity on $T^d$ to odd dimensions generically contain terms of this form. Variation of   (\ref{Lagrangian
5}) with respect to $C_{(p-1)}$ leads to the self-duality constraint
\begin{equation}\label{duality constraint a}
d*G_{(p)}-\mu G_{(p)}=0
\end{equation}
so that $G_{(p)}$ has the number of degrees of freedom one would
expect of a \emph{massive} $p$-form field strength. Applying the
$d*$ operator to this equation produces the equation of motion for
a massive field
\begin{equation}
(\Box-\mu^2)G_{(p)}=0
\end{equation}
The self-duality constraint (\ref{duality constraint a})
implies that $*dC_{(p-1)}-\mu C_{(p-1)}$ is closed so that
locally one  may   introduce a dual potential $\vartheta_{(p-2)}$ such that
\begin{equation}\label{duality relation 2}
*dC_{(p-1)}-\mu C_{(p-1)}=d\vartheta_{(p-2)}
\end{equation}
The gauge invariance of $G_{(p-1)}$ under the transformation $\delta C_{(p-1)}=d\lambda_{(p-2)}$ induces the transformation in the dual
potential $\delta\vartheta_{(p-2)}=-\mu\lambda_{(p-2)}$.

The standard (massless) dualisation techniques do not work in this topologically
massive case as the Lagrangian can not be written solely in terms of $G_{(p)}$. However there is a formalism discussed in
\cite{Betancourt:2004zs} that may be generalised and used to define a dual Lagrangian. Consider the first order Lagrangian
\begin{equation}\label{doubled Lagrangian 3}
{\cal L}_{C,S}=-G_{(p)}\wedge S_{(p-1)}+\frac{1}{2}\mu
G_{(p)}\wedge C_{(p-1)}+\frac{1}{2}*S_{(p-1)}\wedge S_{(p-1)}
\end{equation}
where a $(p-1)$ form field $S_{(p-1)}$ has been introduced. This can be thought of as a doubled formalism as the set of fields has been doubled.
The invariance of this Lagrangian under the gauge transformation $\delta
C_{(p-1)}=d\lambda_{(p-2)}$, up to an irrelevant total derivative, requires that $\delta S_{(p-1)}=0$. Taking the variation of the Lagrangian
(\ref{doubled Lagrangian 3}) with respect to $S_{(p-1)}$ gives the constraint $G_{(p)}=*S_{(p-1)}$, which when substituted back into ${\cal
L}_{C,S}$, gives the topologically massive theory of (\ref{Lagrangian 5}).
 Variation with respect to $C_{(p-1)}$ gives the self duality constraint
$d(*G_{(p)}-\mu C_{(p-1)})=0$ and subsequently the equation of motion $(\Box -\mu^2)G_{(p)}=0$.

Alternatively,  varying ${\cal L}_{C,S}$ with respect to $C_{(p-1)}$ gives the complimentary constraint $dS_{(p-1)}=\mu
G_{(p)}$ which may be written
\begin{equation}\label{stuckelberg constraint}
S_{(p-1)}=\mu C_{(p-1)}+d\vartheta_{(p-2)}
\end{equation}
(\ref{stuckelberg constraint}) may be thought of as a definition
of the massive $S_{(p-1)}$ field in terms of a gauge field
$C_{(p-1)}$ eating $\vartheta_{(p-2)}$. Substituting the
constraint (\ref{stuckelberg constraint}) back into the doubled
Lagrangian gives the dual theory
\begin{equation}\label{dual Lag}
\widetilde{{\cal L}}_{S}=-\frac{1}{2}\mu^{-1} dS_{(p-1)}\wedge
S_{(p-1)}+\frac{1}{2}*S_{(p-1)}\wedge S_{(p-1)}
\end{equation}
The equations of motion from ${\cal L}_C$ and $\widetilde{{\cal L}}_S$ are equivalent so the Lagrangians are classically dual.

In this way, the quadratic  terms of the form (\ref{Lagrangian 5}) for   2-form gauge fields in $D=5$ and 3-form gauge fields in $D=7$
 arising from twisted compactifications with flux can be dualised to  (\ref{dual Lag}), which is the form
 of the quadratic  term for these gauge fields in the Universal Lagrangian in these dimensions given in \cite{deWit:2005hv} and \cite{Samtleben:2005bp}.
 It is to be expected that once  the  gauge group is chosen and the quadratic form of the theories fixed,
 supersymmetry and gauge invariance should determine the theory uniquely.
 As the two theories agree at the quadratic level
and are gauge invariant and supersymmetric, they should be fully  equivalent.
However, the non-linearity of the theory and the need for field redefinitions makes this hard to verify in general.
We shall instead check the full non-linear equivalence in a particular  model simple enough to allow a complete analysis.

\subsection{Example:  Flux Compactifications To Seven-Dimensions}

The Universal Lagrangian in seven dimensions was constructed in \cite{Samtleben:2005bp} along the same lines as the five dimensional case of
\cite{deWit:2005hv} reviewed in section 5.3.
The ungauged theory has an $SL(5,\R)$ symmetry, and this extends to a formal symmetry of the gauged theory if the embedding tensor that  specifies the
embedding of the gauge group in $SL(5,\R)$ also transforms.
The embedding tensor $\Theta_{AB,C}{}^D$   defines the gauge generators ${\cal T}_{AB}$ as
\begin{equation}
{\cal T}_{AB}=\Theta_{AB,C}{}^Dt^C{}_D
\end{equation}
where $t^C{}_D$ are the generators of $SL(5)$ and $A=1,2...5$. It is useful to define the projectors ${\cal Z}^{AB,C}$ and ${\cal Y}_{AB}$ in
terms of the $\mathbf{5}$ and $\mathbf{10}$ representations of the gauge generators ${\cal T}_{AB,C}{}^D$ and ${\cal T}_{AB,CD}{}^{EF}=2{\cal
T}_{AB,[C}{}^{[E}\delta_{D]}{}^{F]}$ respectively, where
\begin{eqnarray}
{\cal T}_{AB,C}{}^D&=&\Theta_{AB,C}{}^D=\delta_{[A}{}^D{\cal Y}_{B]C}-2\varepsilon_{ABCEF}{\cal Z}^{EF,D}
\end{eqnarray}
The theory has the potentials ${\cal A}_{(1)}^{AB}$ in the $\overline{\mathbf{10}}$ of $SL(5)$ and ${\cal B}_{(2)A}$ in the $\mathbf{5}$ of
$SL(5)$. In addition there are self-dual three forms ${\cal S}_{(3)}^A$ in the $\overline{\mathbf{5}}$ representation. The $SL(5)$ and gauge
covariant field strengths for these potentials are
\begin{eqnarray}
{\cal H}_{(2)}^{AB}&=&d{\cal A}^{AB}_{(1)}+\frac{1}{2}g{\cal T}_{CD,EF}{}^{AB}{\cal A}^{CD}_{(1)}\wedge{\cal A}^{EF}_{(1)}+g{\cal Z}^{AB,C}{\cal B}_{(2)C}\nonumber\\
{\cal H}_{(3)A}&=&D{\cal B}_{(2)A}+\varepsilon_{ABCDE}{\cal A}^{BC}_{(1)}\wedge d{\cal A}^{DE}_{(1)}+\frac{2}{3}g\varepsilon_{ABCDE}{\cal
T}_{FG,H}^D{\cal A}^{BC}_{(1)}\wedge{\cal A}^{EH}_{(1)}\wedge{\cal A}^{FG}_{(1)}+g{\cal Y}_{AB}{\cal S}_{(3)}^B\nonumber\\
{\cal H}_{(4)}^A&=&D{\cal S}_{(3)}^A+{\cal F}_{(2)}^{AB}\wedge {\cal B}_{(2)B}+\frac{1}{2}g{\cal Z}^{AB,C}{\cal B}_{(2)B}\wedge {\cal
B}_{(2)C}+\frac{1}{3}\varepsilon_{BCDEF}{\cal A}_{(1)}^{AB}\wedge{\cal
A}_{(1)}^{CD}\wedge d{\cal A}_{(1)}^{EF}\nonumber\\
&&+\frac{1}{6}g\varepsilon_{BCDEF}{\cal T}_{GH,I}{}^E{\cal A}_{(1)}^{AB}\wedge{\cal A}_{(1)}^{CD}\wedge{\cal A}_{(1)}^{GH}\wedge{\cal
A}_{(1)}^{IF}
\end{eqnarray}
where ${\cal F}_{(2)}^{AB}={\cal H}_{(2)}^{AB}-g{\cal Z}^{AB,C}{\cal B}_{(2)C}$.

As an application of the techniques of section 5.4, consider the reduction of eleven-dimensional supergravity on a four-dimensional torus to
seven dimensions. A flux $\cal K$ is introduced as described in section 2. Using the field redefinitions of Appendix B, the Lagrangian of the
reduced theory is (\ref{reduced Lagrangian}) where the Chern-Simons term, given in full in Appendix C, may be written as
\begin{equation}
{\cal L}^{cs}_7=d\widetilde{C}_{(3)}\wedge{\cal Q}_{(3)}-\frac{1}{12}\epsilon^{mnpq}d\widetilde{C}_{(2)m}\wedge
d\widetilde{C}_{(2)n}\wedge\widetilde{C}_{(1)pq}+{\cal L}^{Top}_7
\end{equation}
where $G_{(4)}=d\widetilde{C}_{(3)}+{\cal W}_{(4)}$ and
\begin{equation}
{\cal Q}_{(3)}=\varepsilon^{mnpq}\left(-\frac{1}{6}\widetilde{C}_{(2)m}\wedge dC_{(0)npq}+\frac{1}{8}\widetilde{C}_{(1)mn}\wedge
d\widetilde{C}_{(1)pq}\right)
\end{equation}
is independent of $\widetilde{C}_{(3)}$. ${\cal L}_7^{Top}$ is the topological mass term
\begin{equation}
{\cal L}_7^{Top}=\frac{1}{2}\mu\widetilde{C}_{(3)}\wedge d\widetilde{C}_{(3)}
\end{equation}
and the parameter $\m$ defined by
\begin{equation}
\mu=\frac{1}{24}\epsilon^{mnpq}K_{mnpq}
\end{equation}
is the topological mass of the $\widetilde{C}_{(3)}$ field. As discussed at the end of section 5.4, this mass term prevents the dualisation of
the three form but a dual formulation of the theory may be found following the discussion of section 5.4.2. Consider the first order Lagrangian
\begin{eqnarray}\label{doubled Lagrangian}
{\cal L}_{\widetilde{C},S}&=&-e^{2\alpha\varphi}G_{(4)}\wedge S_{(3)}+d\widetilde{C}_{(3)}\wedge\left({\cal Q}_{(3)}+\frac{1}{2}\mu
\widetilde{C}_{(3)}\right)+\frac{1}{2}*S_{(3)}\wedge S_{(3)}\nonumber\\&&-\frac{1}{12}\epsilon^{mnpq}d\widetilde{C}_{(2)m}\wedge
d\widetilde{C}_{(2)n}\wedge\widetilde{C}_{(1)pq}+{\cal L}'
\end{eqnarray}
generalising that of (\ref{doubled Lagrangian 3}), where ${\cal L}'$ represents all those terms in the ${\cal L}_7$ Lagrangian that neither
depend on $\widetilde{C}_{(3)}$ nor enter into the Chern-Simons term ${\cal L}^{cs}_7$. Taking the variation of ${\cal L}_{\widetilde{C},S}$
with respect to $S_{(3)}$ produces the duality constraint
\begin{equation}\label{duality constraint 1}
S_{(3)}=e^{2\alpha\varphi}*G_{(4)}
\end{equation}
Substituting this back into the first order Lagrangian (\ref{doubled Lagrangian}) gives
 the Lagrangian (\ref{reduced Lagrangian}) for the flux compactification of eleven dimensional supergravity on $T^4$.
If instead, the first order Lagrangian is varied with respect to $\widetilde{C}_{(3)}$ the constraint $d(e^{2\alpha\varphi}S_{(3)}-{\cal
Q}_{(3)}-\mu\widetilde{C}_{(3)})=0$ arises, which may be written as
\begin{equation}\label{duality constraint 2}
e^{2\alpha\varphi}S_{(3)}=H_{(3)}
\end{equation}
where $H_{(3)}=d\vartheta_{(2)}+{\cal Q}_{(3)}+\mu\widetilde{C}_{(3)}$ for some two-form $\vartheta_{(2)}$. Combining the duality constraint
(\ref{duality constraint 2}) with (\ref{duality constraint 1}) gives
\begin{equation}
e^{-4\alpha\varphi}*G_{(4)}=H_{(3)}
\end{equation}
In the case of zero flux $\mu=0$ this duality constraint reduces to that required to produce the $SL(5)$ invariant Lagrangian in the ungauged
theory. Substituting the constraint (\ref{duality constraint 2}) back into the first oredr Lagrangian (\ref{doubled Lagrangian}) gives the dual
theory
\begin{eqnarray}\label{dual Lagrangian}
\widetilde{{\cal L}}_{\widetilde{S}}&=&-\frac{1}{2}\mu\widetilde{S}_{(3)} \wedge\left(d\widetilde{S}_{(3)}+2{\cal W}_{(4)}-2\mu^{-1}d{\cal
Q}_{(3)}\right) +\frac{1}{2}\mu^2e^{-4\alpha\varphi}*\widetilde{S}_{(3)}\wedge
\widetilde{S}_{(3)}\nonumber\\
&&+\frac{1}{2}\mu^{-1}{\cal Q}_{(3)}\wedge d{\cal Q}_{(3)}-\frac{1}{12}\epsilon^{mnpq}d\widetilde{C}_{(2)m}\wedge
d\widetilde{C}_{(2)n}\wedge\widetilde{C}_{(1)pq}+{\cal L}'
\end{eqnarray}
where the three form $\widetilde{S}_{(3)}$ is defined $\widetilde{S}_{(3)}=\mu^{-1}e^{2\alpha\varphi}S_{(3)}$. The gauge variation of the first
two terms in the second line cancel so that this Lagrangian is gauge invariant. Using (\ref{duality constraint 2}) this Lagrangian may be
written in terms of $\widetilde{C}_{(3)}$
\begin{eqnarray}\label{dual Lagrangian}
\widetilde{{\cal L}}&=&-\frac{1}{2}\mu\widetilde{C}_{(3)} \wedge\left(d\widetilde{C}_{(3)}+2{\cal W}_{(4)}\right)
+\frac{1}{2}\mu^2e^{-4\alpha\varphi}*H_{(3)}\wedge
H_{(3)}\nonumber\\
&&-{\cal W}_{(4)}\wedge H_{(3)}-\frac{1}{12}\epsilon^{mnpq}d\widetilde{C}_{(2)m}\wedge d\widetilde{C}_{(2)n}\wedge\widetilde{C}_{(1)pq}+{\cal
L}'
\end{eqnarray}
It is then a straightforward, if laborious, process to show that this Lagrangian is equivalent to the seven dimensional Universal Lagrangian of
\cite{Samtleben:2005bp} with embedding tensor defined by
\begin{equation}
{\cal Z}^{AB,C}=0   \qquad  {\cal Y}_{AB}=\frac{\mu}{2\epsilon_1^2\epsilon_2g}\delta_{(A}^5\delta_{B)}^5
\end{equation}
corresponding to the generalised structure constants
\begin{equation}
{\cal T}_{AB,C}{}^D=\Theta_{AB,C}{}^D=-\frac{\mu}{2\epsilon_1^2\epsilon_2g}\delta_{AB}^{5D}\delta_C^5   \qquad  {\cal
T}_{AB,CD}{}^{EF}=-\frac{\mu}{\epsilon_1^2\epsilon_2g}\delta_{AB}^{5[E|}\delta_{CD}^{5|F]}
\end{equation}
where the constants $\epsilon_1$ and $\epsilon_2$ are determined by the full universal Lagrangian. The potentials of the Universal Lagrangian
are given by the reduced potentials, up to the constant factors $\epsilon_1$ and $\epsilon_2$ as
\begin{eqnarray}
{\cal A}_{(1)}^{5m}&=&\epsilon_1A^m\nonumber\\
\epsilon_2\varepsilon_{mnpq}{\cal A}_{(1)}^{pq}&=&C_{(1)mn}\nonumber\\
{\cal B}_{(2)m}&=&\frac{4\epsilon_1}{\epsilon_2}\left(C_{(2)m}+\frac{1}{2}C_{(1)mn}\wedge A^n\right)\nonumber\\
{\cal S}_{(3)}^5&=&-\frac{4\epsilon_1^2}{\epsilon_2}\left(C_{(3)}-\frac{1}{6}C_{(1)mn}\wedge A^m\wedge A^n\right)
\end{eqnarray}
The field strengths are related by
\begin{eqnarray}
{\cal H}_{(2)}^{5m}&=&\epsilon_1F^m\nonumber\\
\epsilon_2\varepsilon_{mnpq}{\cal H}_{(2)}^{pq}&=&G_{(2)mn}+C_{(0)mnp}F^p\nonumber\\
{\cal H}_{(3)m}&=&\frac{4\epsilon_1}{\epsilon_2}G_{(2)m}\nonumber\\
{\cal H}_{(3)5}&=&-\frac{2}{\epsilon_2^2}\left(H_{(3)}+\frac{1}{6}\varepsilon^{mnpq}C_{(0)mnp}G_{(3)q}\right)\nonumber\\
{\cal H}_{(4)}^5&=&-\frac{4\epsilon_1^2}{\epsilon_2}G_{(4)}
\end{eqnarray}

\subsection{Other dimensions}
In four dimensions the graviphoton field $A^m$ and its dual must be included in the same multiplet to write the ungauged theory in an
$E_{7(7)}$-invariant form. Such a theory in which the isometry symmetry is doubled cannot be given a purely geometric interpretation. If instead
only the $C_{(1)mn}$ fields are doubled then it was shown in \cite{D'Auria:2005er} how the conjectured Lie algebra of this theory could be
embedded in $E_{7(7)}$. In the absence of a the complete $E_{7(7)}$-covariant Universal Lagrangian it is difficult to comment on this
conjecture. However, it is clear that the gauge algebra (\ref{c-field algebra 3}) is a contraction of that presented in \cite{D'Auria:2005er}
and it is therefore plausible that the relation between two gauge theories could be  similar
to the relation between the $CSO(p,q,r)$ and $SO(p+r,q)$ gaugings of maximal supergravity in
four dimensions presented in \cite{Hull:1984yy,Hull:1984qz}.

\section{Non-Geometric Solutions and Duality}

 In this paper we have considered in  detail the Scherk-Schwarz dimensional reduction  with flux of (the bosonic sector of) 11-dimensional supergravity   to any dimension, to define a lower-dimensional gauged supergravity theory.
We expect these to  fit into the general gauged supergravities of \cite{deWit:2004nw,Samtleben:2005bp}, and have checked this in detail in the
case of certain reductions to seven dimensions. We have also addressed the issue of whether these reductions arise from compactifications of
M-theory. In general this is not the case. The Scherk-Schwarz reduction can be thought of as arising from a reduction on a group manifold $G$
followed by a truncation to a finite set of lower-dimensional fields. For this to arise from a compactification with mass gap, it is necessary
that either $G$ is compact, or that there is a discrete left-acting subgroup $\Gamma$ such that $G/\Gamma$ is compact, in which case the
reduction is a truncation of the compactification on $G/\Gamma$, and this can be extended to compactification of M-theory on $G/\Gamma$. This
gives a wide class of explicit flux compactifications of M-theory.

An important feature of the general formulations of gauged supergravity of \cite{deWit:2004nw,deWit:2005hv,deWit:2005ub,Samtleben:2005bp} is that they are covariant under the action of the
$E_{n}$ duality group, and so provide a formalism to discuss the action of duality transformations in such theories. The situation is then
similar to that described in \cite{Odd,Dabholkar:2005ve} for compactifications of the heterotic string. In all of these cases, a conventional
reduction on a torus $T^d$ gives an ungauged supergravity theory with a duality symmetry $U$. Here $U=E_{d+1} $ for reduction of M-theory on
$T^{d+1}$ or type II theory on $T^d$,   and  $U=O(d,d+16)$ for reduction of heterotic
strings on $T^d$.
For the common sector of these theories  has $U=O(d,d)$, and this is also the group for reduction of bosonic strings on $T^d$.
The gauged supergravities are deformations of these theories in which a subgroup $H$ of $U$ is promoted to a local symmetry,
so that the minimal couplings break the  original $U$ symmetry to a subgroup containing $H$. Remarkably,  the remainder of $U$ still has a
natural action; it is no longer a symmetry, but acts on the embedding tensor and gauge coupling constants, so that the mass terms and scalar
potential are changed, as are the minimal couplings. As a result, $U$ acts to take one gauged supergravity to another. In fact they are
equivalent field theories related by a field redefinition, as in \cite{Hull:1984yy,Hull:1984qz}, but the embedding of the gauge group in $U$ is changed to a conjugate one.

However,
the action of $U$ becomes non-trivial if one tries to lift these theories to higher dimensions. This was analysed in detail in \cite{Dabholkar:2005ve}
for the case $U=O(d,d)$. For example, starting with a twisted torus reduction with twist $f_{mn}{}^p$ and $H$-flux $K_{mnp}$ and gauge  algebra
\begin{eqnarray}\label{geometric algebra}
\left[Z_m,Z_n\right]&=&-f_{mn}{}^pZ_p+K_{mnp}X^p\nonumber\\
\left[Z_m,X^n\right]&=&f_{mp}{}^nX^p\nonumber\\
\left[X^m,X^n\right]&=&0
\end{eqnarray}
it was found that some $O(d,d) $ transformations can interchange twist with flux, or mix them together to give a new twisted torus reduction with flux.
However in
 other
cases, an $O(d,d) $ transformation can take a geometric compactification to a
   non-geometric backgrounds such as a T-fold, with T-duality transition functions \cite{Hull:2004in}. This generalises to the case of M-theory reductions
with flux. In some cases the duality might take a  twisted torus reduction with flux to another twisted torus reduction with flux, with
transformation properties for the twist and flux, generalising the Buscher rules, that can be read off from the supergravity. In others it must
give a non-geometric background, such as
 U-folds \cite{Hull:2004in}, with U-duality transition functions.

To see how this works in more detail, recall that the data for the gauged (super)gravity theories arising from (the common sector of)
superstring theory is all contained in the structure constants $t_{AB}{}^C$ for the gauge group, which is a subgroup of $O(d,d)$. As we saw in
section 5.1,  for twisted torus reductions with flux, these are constructed from the twist (the structure constants $f_{mn}{}^p$ of the group
$G$) and the flux $K_{mnp}$. However for a  general subgroup of $O(d,d)$, the Lie algebra will be of the form
\begin{eqnarray}\label{nongeometric algebra}
\left[Z_m,Z_n\right]&=&-f_{mn}{}^pZ_p+K_{mnp}X^p\nonumber\\
\left[Z_m,X^n\right]&=&h_{mp}{}^nX^p+\widetilde{h}_m{}^{np}Z_p\nonumber\\
\left[X^m,X^n\right]&=&\widetilde{f}^{mn}{}_pX^p+\widetilde{K}^{mnp}Z_p
\end{eqnarray}
and these parameterise the general gauged supergravity theory \cite{Dabholkar:2005ve}. The Jacobi identities constrain the structure constants such
that $t_{[AB}{}^Dt_{C]D}{}^E=0$ and the action of the adjoint representation must be trace-free $t_{AB}{}^B=0$.

In a twisted torus reduction, $f_{mn}{}^p$ and $K_{mnp}$ are the twist and flux respectively, suggesting that $\widetilde{f}_m{}^{np}$ and
$\widetilde{K}^{mnp}$ might be thought of as a dual twist and dual flux \cite{Shelton:2005cf,Dabholkar:2005ve}. Under $O(d,d)$ transformations,
$t_{AB}{}^C$ transforms as an $O(d,d)$ tensor, so that in general T-duality mixes $f_{mn}{}^p$, $K_{mnp}$, $h_{mp}{}^n$,
$\widetilde{h}_m{}^{np}$, $\widetilde{f}^{mn}{}_p$ and $\widetilde{K}^{mnp}$. In special cases, a twisted torus reduction specified by
$f_{mn}{}^p$ and $K_{mnp}$ will transform under certain T-dualities to  a new twisted torus reduction specified by some $f'_{mn}{}^p$,
$K'_{mnp}$ so that T-duality mixes the twist and flux to give a new twisted torus reduction of the same form. However,  in general an $O(d,d) $
transformation will lead to a general structure $f_{mn}{}^p$, $K_{mnp}$, $h_{mp}{}^n$, $\widetilde{h}_m{}^{np}$, $\widetilde{f}^{mn}{}_p$ and
$\widetilde{K}^{mnp}$ with dual twist and flux \cite{Shelton:2005cf,Dabholkar:2005ve}. The interpretation of the dual twist and flux was
discussed in \cite{Dabholkar:2005ve} and they typically indicate a non-geometric background. Explicit constructions of such non-geometric
backgrounds were given in \cite{Dabholkar:2005ve}. However, they can also arise from geometric compactifications which are not of twisted torus
type, so that it is misleading to think of $\widetilde{f}_m{}^{np}$ and $\widetilde{K}^{mnp}$ as being intrinsically non-geometric \cite{Dabholkar:2005ve}. In the special case where $K_{mnp}=\widetilde{K}^{mnp}=0$, $h_{mn}{}^p=f_{mn}{}^p$ and $\widetilde{h}_m{}^{np}=\widetilde{f}_m{}^{np}$, the
algebra (\ref{nongeometric algebra}) is a Drinfeld double \cite{Alekseev:1993qs,Klimcik:1995dy}, suggesting a possible relation between the kind
of T-duality considered here and the Poisson-Lie T-duality of \cite{Klimcik:1995dy,Klimcik:1995ux,Klimcik:1996uc}.

The generalisation to the heterotic string is straightforward. The duality group is now $O(d,d+16)$ and    $t _{AB}{}^C$ becomes an $O(d,d+16)$
tensor. From section 5.2, there is an additional flux $M_{mn}{}^a$ and the decomposition of the structure constants $t _{AB}{}^C$ of the general
gauge algebra generalising (\ref{nongeometric algebra}) will have further terms involving the generators $Y_a$.

It is interesting to ask how this extends to M-theory compactifications, and their non-geometric generalisations.
The generator $X^m$ of the heterotic theory, which is  associated with string winding modes, is replaced with the
generator $X^{mn}$ in M-theory,  which might be associated with membrane wrapping modes,  so that one would expect the algebra (\ref{c-field algebra 3}) to be become something like
\begin{eqnarray}\label{nongeometric algebra 2}
\left[Z_m,Z_n\right]&=&-f_{mn}{}^pZ_p+K_{mnpq}\breve{X}^{pq}\nonumber\\
\left[Z_m,\breve{X}^{np}\right]&=&2h_{mq}{}^{[n}\breve{X}^{p]q}+\widetilde{h}_m{}^{npq}Z_q\nonumber\\
\left[\breve{X}^{mn},\breve{X}^{pq}\right]&=&\widetilde{f}_{ts}{}^{mnpq}\breve{X}^{ts}+\widetilde{K}^{mnpqt}Z_t
\end{eqnarray}
where we have used the decomposition of $X^{mn}$ into $\breve{X}^{mn}$ and $\breve{X}^m$ which satisfy $\breve{X}^p=f_{mn}{}^pX^{mn}$ and
$f_{mn}{}^p\breve{X}^{mn}=0$.

As an example of such a reduction, consider the compactification of eleven-dimensional supergravity on $\cX=S^1\times G/\Gamma$ with internal
coordinates $(y^{11},y^m)$ where the flux lies along the circle direction $K_{mnp11}$ and the structure constants of the group $G$ are
$f_{mn}{}^p$. This is a compactification of IIA supergravity on a twisted torus $G/\Gamma$, lifted to M-theory. The Neveu-Schwarz sector of the reduced theory has a
gauge algebra with (\ref{geometric algebra}) as a subalgebra.  Replacing $G/\Gamma$ with a
 non-geometric background  leads to a theory with
gauge algebra containing (\ref{nongeometric algebra}) as a subalgebra. The full algebra of this reduction is then a particular example of the the Lie
algebra (\ref{nongeometric algebra 2}).

A general feature of twisted torus reductions with flux is that the gauge algebra has an abelian subalgebra generated by  $X^m$ or
$\breve{X}^{mn}$ (with $[X,X]=0$). An algebra such as (\ref{nongeometric algebra}) or (\ref{nongeometric algebra 2}) without $[X,X]=0$ can
arise from non-geometric compactifications, but they can also arise from geometric compactifications which are not of Scherk-Schwarz type, such
as the WZW compactifications discussed in \cite{Dabholkar:2005ve} or compactifications on   $S^4$ or $S^7$.

However, this is not quite the whole story. In toroidal compactifications to 6, 5, and 4 dimensions the anti-symmetric tensor gauge fields
$C_{(3)}$, $C_{(2)m}$ and $C_{(1)mn}$ may be dualised to vector fields $\theta_{(1)}$, $\theta_{(1)}{}^m$ and $\theta_{(1)}{}^{mn}$
respectively. These gauge bosons couple to gauge generators $Y$, $Y_m$ and $Y_{mn}$ respectively which in each case may be dualised on the
internal manifold to a vector field  $\theta_{(1) mnnpq}$ coupling to a generator $Y^{mnpqt}$, which might be associated with 5-brane wrapping modes. For twisted torus compactifications with
flux we have seen that the same curvatures and fluxes that allow for a non-abelian gauge symmetry obstruct the dualisation of these fields and
the $C_{(3)}$, $C_{(2)m}$ and $C_{(1)mn}$ potentials remain as massive tensor fields in the gauged supergravity. A particular example is the
case of flux compactifications to seven dimensions, where the $C_{(3)}$ potential cannot be dualised to a two form and instead appears as a
massive field in the gauged supergravity. However, the universal formalism, reviewed in section 5.3, allows for the dual potentials to be
incorporated through a doubling of the degrees of freedom and so this more general construction will contain theories in which  the dual one-forms $\theta_{(1) mnnpq}$
  appear as  gauge fields. The gauge algebra might then be expected to be of a form in which  all of the structure constants are constructed from
 the data $f_{mn}{}^p$ and $K_{mnpq}$, such as
 \begin{eqnarray}\label{dual algebra}
\left[Z_m,Z_n\right]&=&-f_{mn}{}^pZ_p+K_{mnpq}\breve{X}^{pq}\nonumber\\
\left[Z_m,\breve{X}^{np}\right]&=&2f_{mq}{}^{[n}\breve{X}^{p]q}+K_{mqts}Y^{npqts}\nonumber\\
\left[Z_m,Y^{npqts}\right]&=&5f_{ml}{}^{[n}Y^{pqts]l}\nonumber\\
\left[\breve{X}^{mn},\breve{X}^{pq}\right]&=&2f_{ts}{}^{[m}Y^{n]pqts}
\end{eqnarray}
with all other commutators vanishing. As commented on in section 5.6, the geometric reductions considered in this paper produce theories whose
gauge algebra is a contraction of (\ref{dual algebra}) with $[X,X]=0$ and no $Y^{mnpqr}$. For certain dimensions (\ref{dual algebra}) may be
enhanced as in the case \cite{D'Auria:2005er} where a similar algebra appears and includes an additional term
$[Z_m,Z_n]=g\epsilon_{mnpqtsl}Y^{pqtsl}+...$, which may only occur in four dimensions. For the geometric reductions considered in this paper it
was demonstrated that part of the $X^{mn}$ symmetry, given by the projection $\breve{X}^m=f_{np}{}^mX^{np}$, is always broken by any vacuum of
the theory. It seems that a similar statement  holds for the generators $Y^{mnpqt}$ in (\ref{dual algebra}) where the symmetry generated by
$K_{npqt}Y^{mnpqt}=\breve{Y}^m$ is broken by the vacuum. This was certainly the case in \cite{D'Auria:2005er} where $\breve{Y}^m$ could be
identified with $\breve{X}^m$ which ensured that the correct number of gauge degrees of freedom were present in the theory.

All of the gaugings discussed here can be embedded into the universal Lagrangian formalism reviewed in section 5.3. For example in five
dimensions the general gauge algebra is of the form
\begin{eqnarray}\label{universal algebra}
\left[T_i,T_j\right]&=&-t_{ij}{}^kT_k+h_{ij}{}^aT_a+C_{ij}{}^uT_u\nonumber\\
\left[T_i,T_a\right]&=&C_{ia}{}^uT_u\nonumber\\
\left[T_i,T_u\right]&=&D_{iu}{}^vT_v
\end{eqnarray}
where a particular $E_{6(6)}$ basis has been chosen, as described in section 5.3.3 and the structure constants
$t_{ab}{}^c=t_{uv}{}^w=t_{uv}{}^a=t_{ab}{}^u=0$ all vanish by the constraints on the embedding tensor \cite{deWit:2004nw}. This algebra is broken to
\begin{equation}
\left[T_i,T_j\right]=-t_{ij}{}^kT_k+h_{ij}{}^aT_a
\end{equation}
by any vacuum of the theory so that the $T_u$ are always broken and the $T_a$ give a central extension of the algebra generated by $T_i$. A
different choice of $E_{6(6)}$ basis will lead to an equivalent algebra taking a more complicated form. As an example, consider the reduction of
eleven dimensional supergravity on a semi-simple group manifold (with identifications to compactify, if necessary) with flux. As discussed in
sections 4.2 the symmetry generated by the $X^{mn}$ is broken by any vacuum of the theory and we can make the identifications of the generators
$T_i$ in (\ref{universal algebra}) with $Z_m$ and $T_u$ with $X^{mn}$, with no generators $T_a$.
The symmetries generated by $T_u\sim X^{mn}$ are broken, so that the
remaining symmetry algebra is
\begin{equation}
\left[T_i,T_j\right]=-t_{ij}{}^kT_k
\end{equation}

The algebras of the geometric theories we have considered in this paper are described by (\ref{universal algebra}), but this algebra also
contains generalisations of such gauge algebras arising from compactifications whose lift to M-Theory do not admit a geometric interpretation.
These can be systematically studied in a similar way to those discussed in \cite{Dabholkar:2005ve}, and we plan to  return to a discussion of
such non-geometric backgrounds elsewhere.

\newpage

\appendix

\section{Useful Identities for $O^{qt}_{mnp}$ and
$\Pi^{mnp}_{qt}$}

Recall the definitions of the constants $O^{qt}_{mnp}$ and $\Pi^{mnp}_{qt}$ introduced in section 4.2
\begin{eqnarray}
O^{qt}_{mnp}&=&3\delta^q_{[m}f_{np]}{}^t \nonumber\\
\Pi^{mnp}_{qt}&=&\frac{1}{2}\delta^{[m}_qf_t{}^{np]}
\end{eqnarray}
These objects may be shown to satisfy the following useful identities on a twisted torus $\cX=G/\G$ when $G$ is semi-simple.
\begin{eqnarray}
\label{eq:O relations }O^{qt}_{mnp}\Pi^{mnp}_{kl}&=&\delta^{q}_{k}\delta^{t}_{l}+f_{pk}{}^tf_l{}^{qp}
\nonumber\\
O^{[qt]}_{mnp}\Pi^{mnp}_{kl}&=&\delta^{qt}_{kl}-\frac{1}{2}f_{kl}{}^pf_p{}^{qt}
\nonumber\\ O^{qt}_{mnp}\eta_{qt}&=&3f_{mnp} \nonumber\\
\Pi^{mnp}_{qt}\eta^{qt}&=&\frac{1}{2}f^{mnp} \nonumber\\
O^{kt}_{mnp}f_{tl}{}^s+O^{ts}_{mnp}f_{tl}{}^k&=&3O^{ks}_{[mn|t}f_{|p]l}{}^t \nonumber\\ \eta_{mn}&=&\frac{1}{4}O^{ts}_{mpq}O^{pq}_{nts}
\end{eqnarray}
where $\eta_{mn}$ is the Cartan-Killing metric. For reductions involving higher degree forms there are generalisations of these constants. For
example, in the reduction of a $p$-form $\widehat{C}_{(p)}$ the mass term in the reduced Lagrangian for the potential $C_{(i)n_1n_2...n_{p-i}}$
is of the form $(O\cdot C_{(i)})^2$ and the non-linear gauge transformation of the field $C_{(i-1)n_1n_2...n_{p-i+1}}$ with parameter
$\lambda_{(i-1)m_1m_2...m_{p-i}}$ is of the form $\delta_{\lambda}C_{(i-1)}=O\cdot\lambda_{(i-1)}$ where
\begin{equation}
O^{n_1n_2...n_{p-i}}_{m_1m_2...m_{p+1-i}}=\frac{(p-i+1)!}{2(p-i-1)!}
f_{[m_1m_2}{}^{n_1}\delta^{n_2}_{m_3}\delta^{n_3}_{m_4}...\delta^{n_{p-i}}_{m_{p-i+1}]}
\end{equation}

\section{Field Redefinitions for $T^d$ Reductions}

Reduction on $T^d$ corresponds to the case where $f_{mn}{}^p=0$. The Chern-Simons term becomes very complicated under the standard reduction
ansatze. To ease the algebra the following redefinition of the potential can be made, following \cite{Lu:1995yn}
\begin{equation}
\widehat{C}=\widetilde{C}_{(3)}+\widetilde{C}_{(2)m}\wedge\sigma^m
+\frac{1}{2}\widetilde{C}_{(1)mn}\wedge\sigma^m\wedge\sigma^n
+\frac{1}{6}\widetilde{C}_{(0)mnp}\sigma^m\wedge\sigma^n\wedge\sigma^p+\varpi_{(3)}
\end{equation}
where
\begin{eqnarray}
\widetilde{C}_{(3)}&=&C_{(3)}-C_{(2)m}\wedge
A^m+\frac{1}{2}C_{(1)mn}\wedge A^m\wedge
A^n-\frac{1}{6}C_{(0)mnp}A^m\wedge A^n\wedge A^p
\nonumber\\
\widetilde{C}_{(2)m}&=&C_{(2)m}+C_{(1)mn}\wedge
A^n+\frac{1}{2}C_{(0)mnp}A^n\wedge A^p
\nonumber\\
\widetilde{C}_{(1)mn}&=&C_{(1)mn}-C_{(0)mnp}A^p
\nonumber\\
\widetilde{C}_{(0)mnp}&=&C_{(0)mnp}
\end{eqnarray}
The gauge transformations of these potentials are
\begin{eqnarray}\label{new C transformations}
\delta\widetilde{C}_{(3)}&=&\widetilde{C}_{(2)m}\wedge
d\omega^m+d\widetilde{\lambda}_{(2)}\nonumber\\
\delta\widetilde{C}_{(2)m}&=&-\widetilde{C}_{(1)mn}\wedge
d\omega^n+d\widetilde{\lambda}_{(1)m}\nonumber\\
\delta\widetilde{C}_{(1)mn}&=&\widetilde{C}_{(0)mnp}
d\omega^p+d\widetilde{\lambda}_{(0)mn}\nonumber\\
\delta\widetilde{C}_{(0)mnp}&=&-K_{mnpq}\omega^q
\end{eqnarray}
where the gauge parameter $\widehat{\lambda}$ in this basis is defined as
\begin{equation}
\widehat{\lambda}=\widetilde{\lambda}_{(2)}+\widetilde{\lambda}_{(1)m}\wedge\sigma^m
+\frac{1}{2}\widetilde{\lambda}_{(0)mn}\sigma^m\wedge\sigma^n
\end{equation}
For $f_{np}{}^m= 0$ limit the field strengths (\ref{eq:c-field
strengths}) are
\begin{eqnarray}\label{G on torus}
G_{(4)}&=&d\widetilde{C}_{(3)}+d\widetilde{C}_{(2)m}\wedge
A^m+\frac{1}{2}d\widetilde{C}_{(1)mn}\wedge A^m\wedge
A^n+\frac{1}{6}d\widetilde{C}_{(0)mnp}\wedge A^m\wedge A^n\wedge
A^p
\nonumber\\
&&-\frac{1}{24}K_{mnpq}A^m\wedge A^n\wedge A^p\wedge A^q
\nonumber\\
G_{(3)m}&=&d\widetilde{C}_{(2)m}-d\widetilde{C}_{(1)mn}\wedge
A^n+\frac{1}{2}d\widetilde{C}_{(0)mnp}\wedge A^n\wedge
A^p+\frac{1}{6}K_{mnpq}A^n\wedge A^p\wedge A^q
\nonumber\\
G_{(2)mn}&=&d\widetilde{C}_{(1)mn}+d\widetilde{C}_{(0)mnp}\wedge
A^p-\frac{1}{2}K_{mnpq}A^p\wedge A^q
\nonumber\\
G_{(1)mnp}&=&d\widetilde{C}_{(0)mnp}+K_{mnpq}A^q
\nonumber\\
G_{(0)mnpq}&=&-K_{mnpq}
\end{eqnarray}
and satisfy the Bianchi identities
\begin{eqnarray}
dG_{(4)}+G_{(3)m}\wedge F^m&=&0\nonumber\\
dG_{(3)m}+G_{(2)mn}\wedge F^n&=&0\nonumber\\
dG_{(2)mn}+G_{(1)mnp}\wedge F^p&=&0\nonumber\\
dG_{(1)mnp}+G_{(0)mnpq}\wedge F^q&=&0\nonumber\\ dG_{(0)mnpq}&=&0
\end{eqnarray}

\section{Reduction of Chern-Simons Terms}

The eleven dimensional Chern-Simons term with flux on the $\widehat{C}$ field is
\begin{equation}
\mathcal{L}_{11}^{cs}=\frac{1}{6}d\left( \widehat{C}+\varpi_{(3)}\right)\wedge
d\left(\widehat{C}+\varpi_{(3)}\right)\wedge\left(\widehat{C}+\varpi_{(3)}\right)
\end{equation}
This may be   rewritten modulo surface terms as
\begin{equation}
\mathcal{L}_{11}^{cs}=\frac{1}{6}\widehat{G}\wedge \widehat{G}\wedge\widehat{C}+\frac{1}{2}\widehat{G}\wedge\widehat{C}\wedge {\cal
K}+\frac{1}{2}\widehat{C}\wedge {\cal K}\wedge {\cal K} +\frac{1}{6}{\cal K}\wedge {\cal K}\wedge\varpi_{(3)}
\end{equation}
where
\begin{equation}
{\cal K}= \frac{1}{24}K_{mnpq} \s^m\wedge \s^n\wedge \s^p\wedge \s^q
\end{equation}
The last term vanishes if $D>0$ and the third term vanishes for $D>3$.

In the case in which $f_{mn}^p=0$, we can use the field definitions of Appendix B to give the reduction of the Chern-Simons term as
\begin{eqnarray}
\label{eq:chern-simons}{\cal L}^{cs}_{10}&=&\epsilon^m\frac{1}{2}d\widetilde{C}_{(3)}\wedge d\widetilde{C}_{(3)}\wedge \widetilde{C}_{(2)m}
\nonumber\\ {\cal L}^{cs}_{9}&=&\epsilon^{mn}\left(-\frac{1}{4}d\widetilde{C}_{(3)}\wedge d\widetilde{C}_{(3)}\wedge \widetilde{C}_{(1)mn} -
\frac{1}{2}d\widetilde{C}_{(2)m}\wedge d\widetilde{C}_{(2)n}\wedge \widetilde{C}_{(3)}\right) \nonumber\\ {\cal
L}^{cs}_{8}&=&\epsilon_{mnp}\left(\frac{1}{12}d\widetilde{C}_{(3)}\wedge d\widetilde{C}_{(3)}\widetilde{C}_{(0)mnp} -
\frac{1}{6}d\widetilde{C}_{(2)m}\wedge d\widetilde{C}_{(2)n}\wedge \widetilde{C}_{(2)p} - \frac{1}{2}d\widetilde{C}_{(3)}\wedge
d\widetilde{C}_{(m)}\wedge \widetilde{C}_{(1)np}\right)
\nonumber\\
{\cal L}^{cs}_{7}&=&\epsilon^{mnpq}\left(\frac{1}{6}d\widetilde{C}_{(3)}\wedge d\widetilde{C}_{(2)m}\widetilde{C}_{(0)npq} -
\frac{1}{12}d\widetilde{C}_{(2)m}\wedge d\widetilde{C}_{(2)n}\wedge \widetilde{C}_{(1)pq}+\frac{1}{8}d\widetilde{C}_{(1)mn}\wedge
d\widetilde{C}_{(1)pq}\wedge\widetilde{C}_{(3)}\right. \nonumber\\
&&\left.+ \frac{1}{48}K_{mnpq}d\widetilde{C}_{(3)}\wedge \widetilde{C}_{(3)}\right)
\nonumber\\
{\cal L}^{cs}_{6}&=&\epsilon^{mnpqt}\left(\frac{1}{12}d\widetilde{C}_{(3)}\wedge d\widetilde{C}_{(1)mn}\widetilde{C}_{(0)pqt} -
\frac{1}{12}d\widetilde{C}_{(2)m}\wedge d\widetilde{C}_{(2)n} \widetilde{C}_{(0)pqt}+\frac{1}{8}d\widetilde{C}_{(1)mn}\wedge
d\widetilde{C}_{(1)pq}\wedge \widetilde{C}_{(2)t}\right.
\nonumber\\
&&\left.+\frac{1}{24}K_{mnpq}d\widetilde{C}_{(3)}\wedge \widetilde{C}_{(2)t}\right)
\nonumber\\
{\cal L}^{cs}_{5}&=&\epsilon^{mnpqts}\left(\frac{1}{12}d\widetilde{C}_{(2)m}\wedge d\widetilde{C}_{(1)np}
\widetilde{C}_{(0)qts}+\frac{1}{48}d\widetilde{C}_{(1)mn}\wedge d\widetilde{C}_{(1)pq}\wedge \widetilde{C}_{(1)ts}\right. \nonumber\\ &&\left.-
\frac{1}{72}d\widetilde{C}_{(0)mnp}\wedge d\widetilde{C}_{(0)qts}\wedge \widetilde{C}_{(3)}+\frac{1}{48}d\widetilde{C}_{(3)}\wedge
\widetilde{C}_{(1)mn}K_{pqts}+\frac{1}{48}d\widetilde{C}_{(2)m}\wedge \widetilde{C}_{(2)n}K_{pqts}\right)
\nonumber\\
{\cal L}^{cs}_{4}&=&\epsilon^{mnpqtsl}\left(\frac{1}{48}d\widetilde{C}_{(1)mn}\wedge d\widetilde{C}_{(1)pq}\widetilde{C}_{(0)tsl} -
\frac{1}{72}d\widetilde{C}_{(0)mnp}\wedge d\widetilde{C}_{(0)qts}\wedge \widetilde{C}_{(2)l}\right. \nonumber\\ &&\left.+
\frac{1}{144}d\widetilde{C}_{(3)}\widetilde{C}_{(0)mnp}K_{qtsl} - \frac{1}{48}d\widetilde{C}_{(2)m}\wedge \widetilde{C}_{(1)np}K_{qtsl}\right)
\nonumber\\
{\cal L}^{cs}_{3}&=&\epsilon^{mnpqtslj}\left(-\frac{1}{144}d\widetilde{C}_{(0)mnp}\wedge d\widetilde{C}_{(0)qts}\wedge \widetilde{C}_{(1)lj}+
\frac{1}{144}d\widetilde{C}_{(2)m}\widetilde{C}_{(0)npq}K_{tslj}
\right. \nonumber\\
&&\left. + \frac{1}{192}d\widetilde{C}_{(1)mn}\wedge \widetilde{C}_{(1)pq}K_{tslj}+\frac{1}{1152}K_{mnpq}K_{tslj}\widetilde{C}_{(3)}\right)
\end{eqnarray}
The reduction of the full case with both twist and flux is straightforward but leads to more complicated formulae, and these will not be given explicitly here.

\end{document}